\documentclass[review]{elsarticle}
\usepackage{lineno,hyperref}
\usepackage{amsmath,amsthm,amssymb,fancyhdr,enumerate}
\usepackage{graphicx}
\usepackage{subcaption}
\usepackage{epstopdf}
\usepackage{soul}
\usepackage{color}
\usepackage{booktabs}
\usepackage{multicol}
\usepackage{multirow}
\usepackage{float}
\usepackage{amssymb}
\usepackage{algorithm}
\usepackage{algpseudocode}
\graphicspath{{}}
\usepackage[normalem]{ulem}
\usepackage{url}
\theoremstyle{plain}
\newtheorem{theorem}{Theorem}

\newtheorem{lemma}{Lemma}

\theoremstyle{definition}
\newtheorem{definition}{Definition}

\theoremstyle{remark}

\usepackage{enumitem}
\usepackage[gen]{eurosym}

\DeclareMathSymbol{\mlq}{\mathord}{operators}{``}
\DeclareMathSymbol{\mrq}{\mathord}{operators}{`'}

\journal{Transportation Research Part C: Emerging Technologies}

\begin{document}
\begin{frontmatter}

\title{Personalized Dynamic Pricing Policy for Electric Vehicles: Reinforcement learning approach} 
\author[lgd]{Sangjun Bae}
\ead{sj.bae@lgdisplay.com}
\author[chalmers]{Balázs Kulcsár}
\ead{kulcsar@chalmers.se}
\author[chalmers,ntnu]{Sébastien Gros}
\ead{grosse@chalmers.se} 
\address[lgd]{AI/Big Data Optimization Team, LG Display, 10845 Paju-si, South Korea}
\address[chalmers]{Department of Electrical Engineering, Chalmers University of Technology, 412 96 Göteborg, Sweden}
\address[ntnu]{Department of Engineering Cybernetics, Norwegian University of Science and Technology, 7491 Trondheim, Norway}


\begin{abstract}
With the increasing number of fast-electric vehicle charging stations (fast-EVCSs) and the popularization of information technology, electricity price competition between fast-EVCSs is highly expected, in which the utilization of public and/or privacy-preserved information will play a crucial role. Self-interest electric vehicle (EV) users, on the other hand, try to select a fast-EVCS for charging in a way to maximize their utilities based on electricity price, estimated waiting time, and their state of charge. While existing studies have largely focused on finding equilibrium prices, this study proposes a personalized dynamic pricing policy (PeDP) for a fast-EVCS to maximize revenue using a reinforcement learning (RL) approach. We first propose a multiple fast-EVCSs competing simulation environment to model the selfish behavior of EV users using a game-based charging station selection model with a monetary utility function. In the environment, we propose a Q-learning-based PeDP to maximize fast-EVCS' revenue. Through numerical simulations based on the environment: (1) we identify the importance of waiting time in the EV charging market by comparing the classic Bertrand competition model with the proposed PeDP for fast-EVCSs (from the system perspective); (2) we evaluate the performance of the proposed PeDP and analyze the effects of the information on the policy (from the service provider perspective); and (3) it can be seen that privacy-preserved information sharing can be misused by artificial intelligence-based PeDP in a certain situation in the EV charging market (from the customer perspective).
\end{abstract}
\end{frontmatter}
\begin{table}[!ht]
\scriptsize
\caption{Nomenclature}
\label{Table_Notation}
\centering
\begin{tabular}{p{0.13\linewidth}p{0.57\linewidth}}
\hline
\textbf{Symbol} & \textbf{Description} \\
\hline
\hline
$\mathcal{B}$	 	    & a set of $n_b$ EV users \\
$b_i$		            & the $i$-th EV user \\
$\mathcal{E}$ 		    & a set of $n_e$ fast-EVCSs \\
$e_j$			  	    & the $j$-th fast-EVCS \\
$c_j$			  	    & the number of total charging slots of fast-EVCS $e_j$ \\
$c^{\text{occ,t}}_j$	& the number of occupied charging slots of fast-EVCS $e_j$ at time $t$ \\
$p^t_{(i,j)}$ 		    & the personalized electricity price to EV user $b_i$ offered by fast-EVCS $e_j$ at time $t$ \\
$p^t_{(\cdot,j)}$ 		& the public electricity price broadcasted by fast-EVCS $e_j$ at time $t$ \\
$SOC^t_i$		        & the state of charge of EV user $b_i$ at time $t$ \\
$eSOC_{(i,j)}$		    & the estimated state of charge of EV user $b_i$ arrival at the fast-EVCS $e_j$ \\
${SOC}^\text{full}$     & the full charged SOC \\
$d_{i,j}$               & the binary decision variable indicating whether or not EV user $b_i$ charges at fast-EVCS $e_j$ \\
$\check{s}_{(i,j)}$     & the binary variable indicating whether $b_i$ can travel to $e_j$  \\ 
$s(\cdot)$              & the strategy of EV user $b_i$ corresponds to its fast-EVCS selection \\ 
$t^\text{ch}_{(i,j)}$   & the actual charging time for EV user $b_i$ at fast-EVCS $e_j$ \\
$w^t_{(i,j)}$           & the expected waiting time of EV user $b_i$ at fast-EVCS $e_j$ at time $t$ \\ 
$u^t_{(i,j)}$           & EV user $b_i$'s monetary utility function for fast-EVCS $e_j$ at time $t$  \\ 
B                       & the capacity of the battery \\
$c_\text{f}$            & the fixed cost per charging \\
$k_\text{s}$            & the SOC weighting for the monetary utility function  \\ 
$k_\text{t}$            & the waiting time weighting \\
$k_\text{eff}$          & the energy efficiency of EVs \\
$k_\text{ct}$           & the maximum charging time per EV \\
$p^\text{vtt}$          & the value of travel time \\
$g_j$                   & the group of EV users who selects the fast-EVCS $e_j$ \\
$\mathbf{P}$            & a Nash stable partition  \\
\hline
\end{tabular}
\end{table}

\section{Introduction}
With increasing concerns about environmental issues around the world, electric vehicles (EVs) are emerging as a sustainable and environmentally friendly alternative to conventional combustion vehicles. In the Nordic region, especially, approximately 250,000 EVs were registered by the end of 2017, which made up roughly 8$\%$ of the global total of EVs in 2016 \cite{Cazz2018}.

One limitation of the EVs is their low battery capacity and their long charging time, which require an extensive charging station network. Many studies have been conducted on deploying public fast-electric vehicle charging stations (fast-EVCSs), especially on highways (motorways/freeways) \cite{pereira2022short,DABIRI2015585,7539538,CSIKOS2017429,CSIKOS2017120,pereira2022parameter}, as well as home-based charging facilities \cite{Cazz2018, Chen2016, Lee2017a, Zhang2017a, Yldz2019, Xu2020}. For example, the electric mobility-related initial funding of the European Commission mostly invested in research on developing a cost-efficient fast-charging network, followed by installing fast-charging stations along highways. In Norway, a company established a scheme to deploy a public fast-EVCS at least every 50km on the highway network, and except for the far northern part of Norway, all highways were served by the end of 2017 \cite{Lorentzen2017}. In Sweden, 'Klimatklivet' (Climate Leap) is making huge investments in charging stations (over 12,000 charging points, including fast-EVCSs along several key highways). In such a situation, the electricity price competition among the fast-EVCSs will become a practical and important issue.

From the EV user perspective, on the other hand, EV users might selfishly select one fast-EVCS based on the information obtained. Through the internet or mobile applications, the EV users are able to identify nearby fast-EVCSs and their status, such as electricity prices, locations, available charging slots, etc. Therefore, the interaction between the fast-EVCSs, between the EV users, and between the fast-EVCSs and the EV users will dynamically influence each other's decision-making.

\subsection{Decision-making for pricing problem}
There has been a considerable amount of research recently to shed some insight into the decision-making problem. The following studies have been focusing on power grid stability and incorporating pricing from a network balancing perspective. In order to achieve this, we modeled the EV user's behavior with different levels of accuracy. In the following, we hence outline them as \emph{price-responsive} algorithms with the following attributes: i) optimization/solution method; ii) detailedness of EV models. Most of the methods simplify the EV user's decision and propose multilayer, centralized, or distributed (and often game-theoretic) solutions.


In \cite{Yuan2017}, the authors considered both the charging station pricing problem and the charging station selection problem using a two-stage Stackelberg model. The authors assumed the same state of charge for the EVs (i.e., same charging cost, same driving range). In \cite{Dong2018}, the authors proposed a fast-EVCS pricing strategy for the voltage control of electricity distribution networks using a double-layer optimization model to minimize the total voltage magnitude deviation of distribution networks without loss of income. The authors could not consider the user's willingness, which determines the demand for the electricity distribution network. 
In \cite{Wang2018a}, the authors proposed the joint admission and pricing approach to maximize a charging station's profit, which also allows them to find a trade-off between the revenue of the charging station and the EV users' satisfaction (waiting time). The authors assumed that a charging station makes admission decisions for homogeneous EV users' charging (i.e., the charging station can refuse some EV users to charge to maximize their profit).
In \cite{Hu2016}, pricing mechanisms were proposed under non-cooperative and cooperative scenarios, respectively, for the evenly distributed charging load with the assumption that the charging decisions were made by an aggregator.
In \cite{Bayram2015}, a single-leader-multi-follower Stackelberg game is proposed, in which the charging network operator acts as the leader and the EV customers are the followers. The leader optimizes the prices of charging stations, and each follower makes a selection between the nearest charging station and a less busy one. 
For maintaining power grid stability, in \cite{Ban2012}, the authors proposed the customer allocation algorithm for maximizing their utility based on the waiting time.
In \cite{Escudero-Garzas2012}, the authors studied the pricing of charging stations using a non-cooperative oligopoly game with assumptions such as homogeneous charging stations, homogeneous electric vehicles, and no road system constraints. In \cite{Lee2015}, the authors resolved the charging station selection problem based on the electricity price offered by the charging station, its distance from the charging station, and EV users' preferences for a certain charging station. They assumed that the demand for each charging station increases as its charging price decreases. In \cite{Moghaddam2019}, the authors resolved the overlapping issues that occur at peak time by dynamically adjusting the prices of the charging stations. Here, it is assumed that the charging stations cooperatively determine the prices. In \cite{LATINOPOULOS2017175}, the authors conducted a survey to understand how electric vehicle drivers respond to uncertain future prices when they charge their vehicles away from home. The motivation of the authors is that dynamic pricing of electricity can help operators spread the demand for electricity to avoid costly infrastructure investments. The results suggest that electric vehicle drivers are generally risk-averse when it comes to uncertain pricing, and there is a non-linearity in their choices. In \cite{Wu2017}, the authors proposed a centralized EV charging-control model that schedules the charging of EVs. Through the case study, they showed the benefits of the model in reducing charging costs, negative impacts on the distribution system, and unserved EV charging demand compared to simpler heuristics.

Finally, in \cite{YANG2021103186}, the authors proposed dynamic \emph{demand-responsive} price adjustment schemes based on charging station queue lengths. The results showed that these schemes were effective in balancing charging demand across fast-charging stations, reducing average waiting time, increasing charging station revenue, and improving social welfare.
As indicated in \cite{YANG2021103186}, using more sophisticated charging and EV user models in combination with scalable price shaping algorithms is of capital importance. In this line of thought, data-dependable methods may bring new aspects to naturally integrating model parameter changes into price decision algorithms.

\subsection{Simulation environment}
Electric vehicle charging stations are categorized as follows:
1) Level-1 chargers, which are often used at home, use 120 V alternating current (AC). Unlike other chargers, level-1 chargers do not require the installation of any additional equipment; 2) Level-2 chargers, which are often used for both residential and commercial EVCSs, use a 240 V (for residential) or 208 V (for commercial) plug; and 3) Level-3 chargers, also known as direct current (DC) fast chargers, are only used in commercial and industrial applications.

\noindent Most of the existing studies focused on level-1 or level-2 charging stations due to the low penetration rate of fast-EVCSs (level-3 charging stations) and their limited data. Also, in order to solve the equilibrium problem analytically, the existing studies disregarded the dynamic environment and practical factors in it such as the travel range of the EV users, different charging times, the number of charging slots, etc. In practice, however, the dynamical situation and these factors are expected to play a critical role in EV users' charging station selections due to the high penetration rate of internet applications. {In addition, recently, sharing information about vehicles and customers has become an issue \cite{Mourad2019, Wang2019a, Buckley2018, Kyriakidis2015, Fagnant2015, Sun2013}.} For example, personal data protection is a big issue in optimizing the sharing mobility operation \cite{Mourad2019}. As another example, in the ride-sourcing system, information sharing and disclosure of customers and sources (vehicles) is an important issue \cite{Wang2019a}. Hence, it is necessary to analyze and investigate the effects of this personal information on the EV charging market, such as revenue improvement for service providers, misuse of privacy-preserved information, etc. Especially, privacy protection is now a great concern of persons, industries, governments, etc. in the new era of cloud and big data. There are research projects focusing on privacy-preserved information sharing \cite{ismagilova2020security, huang2017secure, vaidya2021privacy, BAE2021}. In this study, we assume that the personal information used is protected.

\begin{figure*}[!ht]
    \centering
    \includegraphics[width=0.99\textwidth]{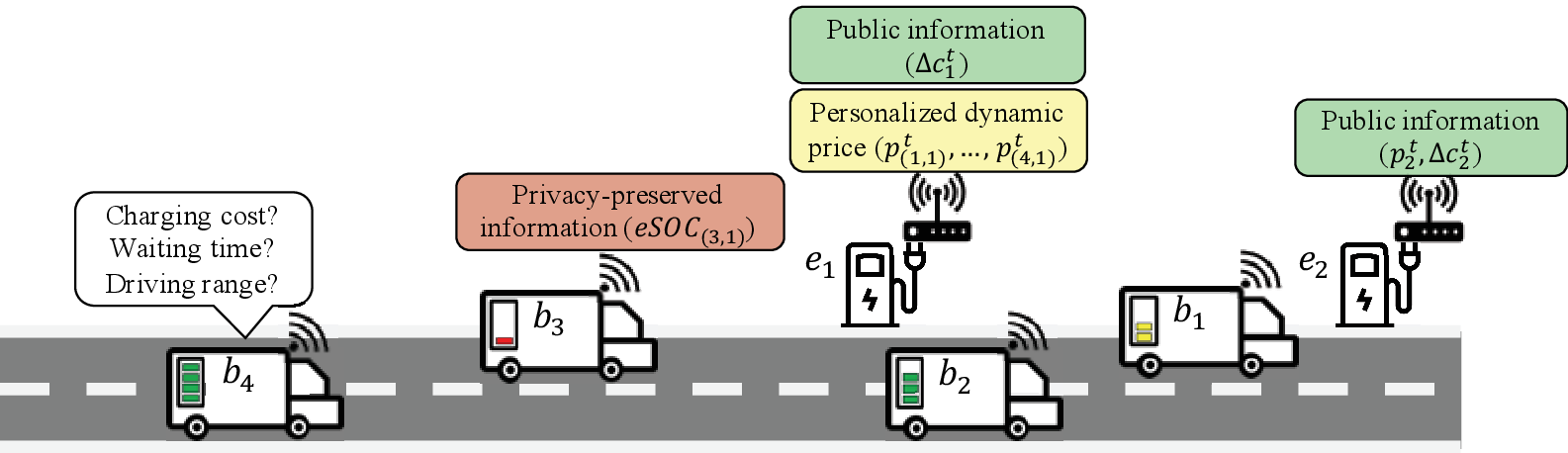}
    \hfil
    \caption{Conceptual image for our system model (highway scenario). Each fast-EVCS adjusts its charging price based on public (green box) and/or privacy-preserved (red box) information. Each selfish EV user selects a fast-EVCS for charging based on the public and/or personalized (yellow box) information.}
    \label{fig:concept}
\end{figure*}

\vspace{1em}
The main aim of this study is to investigate this highly interdependent and complex decision-making problem, in which all stakeholders make self-interested decisions. In this way, we will focus on \emph{demand reponsive}  pricing algorithms (in contrast to the price-responsible approaches presented in Section 1.1.) with detailed EV models. We propose a 1-dimensional highway simulation environment that is dynamically interacting and includes more practical factors, such as estimated state of charge (driving distance to empty), the number of available charging slots, the self-interested decision-making of all stakeholders, non-cooperative decision-making, and a monetary utility function, than the existing studies \cite{Yuan2017, Wang2018a, Bayram2015, Moghaddam2019}. Then, we propose a personalized dynamic pricing policy (PeDP) for fast-EVCSs using a reinforcement learning (RL) approach \cite{Sutton2018}, since it is highly expected that fast-EVCSs might apply pricing policies using artificial intelligence (AI) to maximize their revenue because the charging environment is highly coupled with many agents’ decisions, competitors pricing policies, government policies, etc., which are hard to expect/predict. In such an environment, it is hard to utilize rule-based approaches or optimization-based approaches. It is assumed that the EV user’s privacy is already protected and available based on a technique discussed in \cite{ismagilova2020security, huang2017secure, vaidya2021privacy}. As shown in Figure \ref{fig:concept}, EV users transfer information on their estimated state of charge (eSOC, privacy-preserved information, i.e., the information is only shared with the fast-EVCS) to the fast-EVCSs while they are on the highway. An AI fast-EVCS dynamically adjusts the personalized electricity price (\textit{action}) that is offered to the corresponding EV user, in which the price is affected by public and/or privacy-preserved information from the competing fast-EVCSs and the EV users, respectively (\textit{environment}). Each self-interested EV user updates its charging station selections based on the offered personalized electricity price and other given information (eSOC, available charging slots), in which a game-theoretical approach is applied to model the selfish behavior of the EV users. When the EV users arrive at the AI fast-EVCS for charging, the AI fast-EVCS charges an electricity bill to the EV users (\textit{reward}). We conduct numerical simulations in the proposed environment with fast-EVCSs and dynamically arriving EV users in order to 1) figure out the systemic differences regarding the waiting time between the petrol stations and fast-EVCSs, the proposed PeDP is compared with the classic Bertrand competition model (from the system perspective); 2) evaluate the performance of the proposed PeDP and analyze the importance of the information on the policy (from the fast-EVCS perspective); and 3) show that sharing the privacy-preserved information can be misused by AI-based PeDP in a certain situation in the EV charging market (from the EV user perspective).

Our main contributions are summarized as follows.
\begin{itemize}
    \item \textit{A system model:} The proposed environment considers practical factors such as heterogeneous EVs (e.g., EV-dependent charging time, SOC (travel range)) and dynamically changing capability of the fast-EVCSs, where the parameters for the EVs and fast-EVCSs are based on the public performance data \cite{teslasemi2020} (Q3-1). We model the charging station selection problem as \textit{an anonymous hedonic game} \cite{Darmann2012a, Darmann2018}, which is suitable for modeling the situation where the EV users greedily select fast-EVCSs based on their utility function. Here, we propose a monetary utility function for EV users by transforming the expected waiting time into money using the value of travel time (VTT) \cite{Gunnel2023}. To the best of our knowledge, this is the first work that proposes a flexible and extensible EV charging market simulation environment for the personalized dynamic pricing problem.
    
    \item \textit{Personalized dynamic pricing policy:} Due to the dynamical and highly interdependent interactions between EV users and fast-EVCSs in the proposed environment, it might not be easy to apply the existing approaches, such as the Stackelberg game model \cite{Yuan2017a}. We propose the PeDP using the Q-learning algorithm that interacts with the proposed environment. Different from existing pricing literature, the proposed approach finds the PeDP of the fast-EVCS according to the dynamic environment instead of finding the price equilibrium in the static environment.
    
\end{itemize}

Finally, the above-mentioned methodological solutions have been tested in a numerical simulation context, demonstrating the benefits of dynamic price shaping. As such, we evaluate the PeDP and investigate the importance of information usage in the EV charging market. Firstly, we show the importance of the waiting time in the EV charging market by comparing the PeDP and the classic Bertrand model. Secondly, we evaluate the performance of the PeDP and investigate the effect of the public and/or privacy-preserved information by varying the observable states for the AI fast-EVCS from the service provider perspective. Finally, we present a privacy-preserved information misuse case by the AI-based PeDP in a certain geographical condition.

The remainder of this paper is organized as follows: Section \ref{sec:systemModel} presents the EV charging market simulation environment, which is used as the environment for the model-free reinforcement learning approach. In Section \ref{sec:CSSG}, we propose a charging station selection game (CSSG) model that is a key part of the environment. In Section \ref{sec:DPpolicy}, we propose a PeDP algorithm using a model-free RL approach. Numerical simulations are provided in Section \ref{sec:simulation}. Finally, conclusions and future work are presented in Section \ref{sec:conclusion}.

\section{System Model}
\label{sec:systemModel}

In the EV charging market simulation environment as shown in Figure \ref{fig:concept}, we consider fast-EVCSs and multiple EVs along the highway. There exists a set of EV users $\mathcal{B} = \{ b_1, b_2, \ldots, b_{n_b} \}$ and their state of charge $ SOC^t_{{i}}, \; \forall b_i \in \mathcal{B}$ at time $t$ and a set of fast-EVCSs $\mathcal{E} = \{ e_1, e_2,  \ldots, e_{n_e} \}$ with the number of charging slots $c_j$ and the number of occupied charging slots $c^{\text{occ},t}_j, \; \forall e_j \in \mathcal{E}$ at time $t$ ($c^{\text{occ},t}_j \leq c_j $). The fast-EVCS $e_j$ who adopts the PeDP offers personalized electricity price $p^t_{(i,j)}$ to the corresponding EV user $b_i$ at time $t$. Note that the personalized electricity price information is shared neither with competing fast-EVCSs nor with other EV users. The fast-EVCS $e_j$, which adopts the public dynamic pricing policy (PuDP), broadcasts public electricity price $p^t_{(\cdot,j)}$ at time $t$. Then, EV users charge at one of the fast-EVCSs along the highway. The environment shown in Figure \ref{fig:concept} works by interacting with the following two stages: 
\begin{itemize}
    \item Each fast-EVCS adjusts the personalized or public electricity price in accordance with the PeDP or PuDP, respectively, which is based on both the public information (the number of available charging slots, electricity prices of competitors), and the privacy-preserved information (estimated SOCs of anonymous EV users). The fast-EVCSs announce their current status that is somehow required to be broadcast, such as the number of available charging slots, locations, and public electricity price (if the price is not the personalized electricity price).
    \item Each EV user greedily selects a fast-EVCS for charging in a way to maximize its utility using the public information from the fast-EVCSs. Note that every EV user can access public information such as public electricity price (personalized electricity price, if the fast-EVCS adopts the PeDP), the number of available charging slots, and location of each fast-EVCS.
\end{itemize}

In the environment, the following assumptions are made: (1) for simplification, in the environment, there is only one entrance and exit (i.e., the EV users leave the highway at the same exit), see \cite{Yuan2017}; (2) the EV users can only drive in one direction (left to right in the figure) on the multi-lane highway, see \cite{Yuan2017}; (3) for excluding unfair markets such as collusion, monopoly market, etc., it is assumed that each fast-EVCS is owned by a different service provider and they are competing with each other by adjusting only the public or personalized electricity price, see \cite{Yuan2017}; (4) the EV users who have the same vehicle specification behave selfishly with regard to charging station selections in a way to maximize their own utility, see \cite{Yuan2017, Dong2018, Hu2016}; (5) due to the limitation of the lithium-ion battery capacity, charging is necessary, but EV users will charge only once before leaving the highway without exception, see \cite{Yuan2017, Bayram2015}; (6) for simplicity and considering a rational EV charging strategy, EV users charge up to 80\% of their full SOC; (7) fast-EVCSs provide charging service on a first come first served basis (i.e., no reservation system is considered in this study), see \cite{Dong2018, Ban2012}; and (9) as it is “personalized price”, we assume that no one will share the price with others.

\section{Charging Station Selection Game (CSSG)}
\label{sec:CSSG}
In this section, we propose the charging station selection game (CSSG) model for selfish EV users who are willing to maximize their utilities, in which a monetary utility function is also proposed. Then, we introduce a decentralized algorithm to find the equilibrium of the CSSG model, which is used for the demand of the fast-EVCSs.

The CSSG is defined as follows:
\begin{itemize}
    \item Players: The EV users
    \item Strategies: The strategy of an EV user $b_i$ corresponds to its fast-EVCS selection $s_i(\cdot) \in \mathcal{E}$, where $s_i$ is a function of estimated SOC (eSOC, i.e., SOC at arrival), public or personalized electricity price and estimated waiting time.
\end{itemize}

\subsection{A monetary utility function}
\label{sec:utilityFunction}
Due to the lack of data or accurate models for the EV charging market, we propose a new monetary utility function rather than using coefficient-dependent price elasticity models \cite{Bhattacharya2016,Li2011b,Bitar2013,Gharesifard2013}. EV users selfishly determine a fast-EVCS for charging in a way that minimizes the cost of charging. To represent the EV user's perspective, in this study, we propose a monetary utility function, $u^t_{(i,j)}$, for EV user $b_i$ corresponding to $e_j$ at time $t$. The monetary utility function can be defined as follows:
\begin{equation}
    \label{monetaryUtility}
    u^t_{(i,j)} = -\Big( \frac{\check{s}_{(i,j)}(c_\text{f} + p^t_{(i,j)}\Delta{SOC_{(i,j)}}0.8\text{B})}{\Delta{SOC_{(i,j)}}} + p_\text{vtt}w^t_{(i,j)} \Big)
\end{equation}
\noindent where $\check{s}_{(i,j)}$ is a binary variable that indicates whether EV user $b_i$ can travel to $e_j$, $c_\text{f}$ is a fixed cost\footnote{This is the cost that EV users pay each time they visit the charging station, which is inversely proportional to the SOC level. In other words, it is more efficient to charge up to 70\% of SOC than to charge up to 50\% of SOC when an EV user visits a fast-EVCS. Also, it is divided by $\Delta{SOC_{(i,j)}}$ to consider the EV user's saving behavior.} (5 [euro/charging]), and $\text{B}$ is the capacity of the battery (800 [kWh]). Total charging amount is defined as $\Delta{SOC_{(i,j)}}=0.8{SOC}^\text{full}-{eSOC}_{(i,j)}$, where ${eSOC}_{(i,j)}$ is estimated SOC of the EV $b_i$ arrival at the fast-EVCS $e_j$ \cite{teslasemi2020}.

\noindent\textbf{\textit{- Electricity price:}} Electricity price is one of the most dominant factors that influence the EV users' charging station selections. For simplicity, we assume that the minimum and the maximum electricity price are determined by wholesale electricity price and government restrictions, respectively. The lower and upper bounds for public and personalized electricity price $p^t_{(i,j)}$ is as follow:
\begin{equation}
\label{eq:PriceBound_PDP}
    p_{\min} \leq p^t_{(i,j)} \leq p_{\max}, \quad \forall b_i\in \mathcal{B}, \forall e_j \in \mathcal{E},
\end{equation}

\noindent where we assume that the $p_{\min}=0.35 \; \text{[euro/kWh]}$ and $p_{\max}=0.55 \; \text{[euro/kWh]}$ are applied to all fast-EVCSs. While the lower and upper price bounds are public, the personalized electricity price is not public. The boundary prices are approximately determined based on ``Price ranges for a selection of charging practices in the Nordic region'' on page 45 in \cite{Cazz2018}.

\noindent\textbf{\textit{- Estimated state of charge (eSOC):}} Due to the long charging time and the limited number of fast-EVCSs in public places, the SOC plays a critical role in its charging station selection. The EVs' SOCs determine the travel range as well as the income of the fast-EVCS. For example, a short travel range might constrain the charging station selection options. In addition, the $eSOC_{(i,j)}$ affects the term $c_{\text{f}}/{\Delta SOC_{(i,j)}}$, which determines drivers who want to minimize the fixed cost.
\begin{figure}[!ht]
    \centering
    \includegraphics[width=0.5\linewidth]{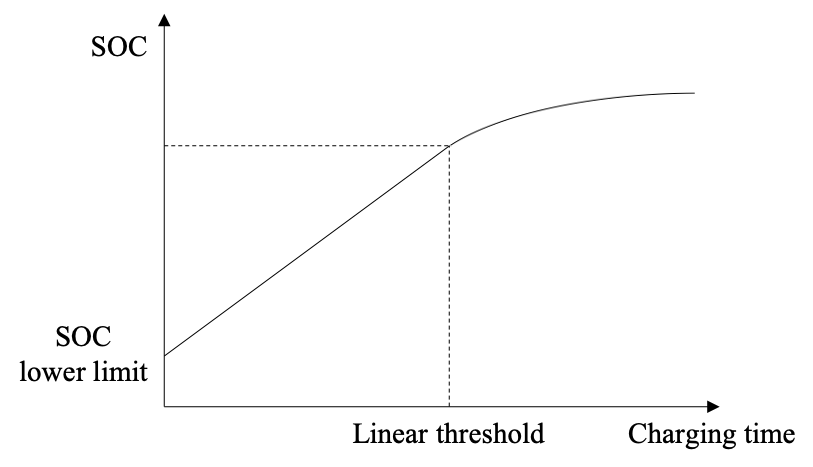}
    \hfil
    \caption{Battery SOC obtained versus charging time spent on a Li-ion battery. Typically, the last phase of the charging curve indicates that the available battery SOC has a nonlinear dependence on the charging time spent \cite{Wang2016}}
    \label{fig:SOCvsTime}
\end{figure}
The time taken to complete a full charge is longer compared to the initial phase of charging, which is between 0\% and 80\%, approximately as shown in Figure \ref{fig:SOCvsTime} \cite{Wang2016}. For considering a rational charging strategy due to these charging characteristics, it is assumed that EV users charge up to 80\% of their full SOC. Also, it is assumed that there are no uncertainties in the travel range of the EVs. Therefore, we define the estimated state of charge $eSOC_{(i,j)}$ of EV user $b_i$ arrival at fast-EVCS $e_j$ as follows:

\begin{equation}
\begin{aligned}
    eSOC_{(i,j)} = SOC_{(i,j)} - & \frac{1}{k_{\text{eff}}} \cdot dist(i,j), \\
    & \quad \forall b_i \in \mathcal{B}, e_j \in \mathcal{E}
\end{aligned}
\end{equation}
where $dist(i,j)$ and $k_{\text{eff}}$ are the euclidean distance between EV user $b_i$ and fast-EVCS $e_j$, and energy efficiency that is assumed as a constant ($k_{\text{eff}}=800$[km/SOC]) in this study, respectively.
Note that it is assumed that the speed of each EV is constant ($100$[km/h]), and factors that impact the energy efficiency, such as temperatures, aging effects, etc., are constant.

\noindent\textbf{\textit{- Expected waiting time:}} As EV users have to share energy resources from the power grid, which has a finite set of charging slots, the expected waiting time may be considerable. Therefore, the expected waiting time will greatly influence the decision-making of the EV users. For example, the EV users will visit a fast-EVCS with a short expected waiting time because other fast-EVCSs do not have enough available charging slots. Unlike the utility functions of the existing studies, which depend on weighting factors \cite{Bhattacharya2016,Li2011b,Bitar2013,Gharesifard2013}, we convert the waiting time to money by using the value of travel time (VTT), $p_\text{vtt}$, which refers to the cost of time spent on transport \cite{Gunnel2023}. $p_\text{vtt}$ is multiplied by the expected waiting time $w^t_{(i,j)}$, which is defined as follows:
\begin{equation}
\label{eq:utility:waiting}
\begin{split}
w^{t}_{(i,j)} & =\left\{\begin{array}{cl}
0,   & \text{if } \Delta{c^t_j} > 0 \; \& \; \Delta{c^t_j} \geq \overline{\overline{g_j}} \\
k_\text{ch}\frac{\overline{\overline{g_j}}}{c_j+\Delta{c^t_j}},   & \text{else if } \Delta{c^t_j} > 0 \; \& \; \Delta{c^t_j} < \overline{\overline{g_j}} \\
k_\text{ch}\frac{|\Delta{c^t_j}| + \overline{\overline{g_j}}}{c_j},   & \text{otherwise} \\
\end{array}\right. \\ &  \qquad \qquad \qquad , \forall b_i \in \mathcal{B}, \forall e_j \in \mathcal{E}, 
\end{split}
\end{equation}
\noindent where $k_\text{ch}$ is a constant value for the charging time from 0.2 to 0.8\footnote{Maintaining the SOC between 0.2 and 0.8 is to keep the battery health the SOC level, which is well-known information. In this study, we assume that the EV users try to maintain the range.}, which takes 3/10 hours \cite{teslasemi2020}. $\overline{\overline{g_j}}$ is the number of tentative decisions of anonymous EV users who want to charge at $e_j$. $c_j$ is the number of total charging slots of $e_j$. $\Delta{c_j^t}$ is the number of available charging slots at $e_j$ at time $t$, defined as $\Delta{c^t_j} = c_j-c^{\text{occ},t}_j$. $c^{\text{occ},t}_j$ is the number of occupied charging slots of $e_j$ at time $t$. The first condition of Eq.(\ref{eq:utility:waiting}) is when there is more than one available charging slot and the number of tentative EV users is less than or equal to the number of available charging slots at $e_j$ at time $t$. The second condition is when there is more than one available charging slot and the number of tentative EV users is greater than the number of available charging slots at $e_j$ at time $t$. The third condition is when there is no available charging slot and there is more than one EV user who tentatively wants to visit at $e_j$ at time $t$. The expected waiting time is converted to money by multiplying the VTT, $p_\text{vtt}$, which is 29 [euro/hour] \cite{Gunnel2023}.

\subsection{Algorithm for CSSG}

The CSSG is solved using a decentralized algorithm \cite{Jang2018g}. Each time step, which is the same update frequency as the fast-EVCS status update in this study, each EV user makes a decision to maximize their monetary utility. EV users' competitive and strategic behaviors can be modeled from game-theoretic perspectives. Here, it is assumed that each EV user is able to access information from the fast-EVCSs, such as locations, electricity prices, and the number of available charging slots. If every EV user is satisfied with their charging station selection, it can be considered that they are mutual-exclusively grouped\footnote{For the sake of convenience, we use the term ``group'' that only indicates the EV users who select the same fast-EVCS.} for each of the fast-EVCSs. This status can be described as the concept called \emph{Nash stable partition}, which is denoted by $\mathbf{P}$. A Nash stable partition always exists and is computed by Algorithm \ref{algorithm:GRAPE}. More details of the algorithm are given in \ref{App:algorithm}.

\section{Personalized Dynamic Pricing Policy Using RL}
\label{sec:DPpolicy}
In this section, we first introduce the PeDP problem to maximize the fast-EVCS' revenue in the proposed environment presented in the previous section. Then, we propose a reinforcement learning approach so that the fast-EVCS will learn its PeDP in the environment.

\subsection{Personalized Dynamic Pricing Problem}

The fast-EVCSs compete with each other to maximize their revenue by adjusting their personalized electricity prices dynamically. The revenue of each fast-EVCSs is determined by the number of visited EVs and the amount of electricity they charge. Here, we assume that other operating costs of charging stations are fixed and not included in mitigating other effects. Therefore, the goal of the fast-EVCS is to perform personalized dynamic pricing that maximizes its revenue as follows:
\begin{equation}
    \label{P:pricing}
    \begin{aligned}
         & \max_{\{ p^t_{(i,{j})} \}}
         & & \sum\limits_{t=1}^{\infty}{\sum\limits_{\forall b_i \in g_{{j}}} {p^t_{(i,{j})}(eSOC_{(i,{j})})} d_{(i,{j})}} \\
         & \text{subject to}
         & & \mathbf{P} \text{ is } \textit{a Nash stable partition}\\
         &
         & & p_{\min} \leq p^t_{(i,{j})} \leq p_{\max} \\
         & 
         & & d_{(i,{j})} \in \{ 0,1 \}\\
         & 
         & & \qquad \qquad , \forall b_i \in \mathcal{B}
    \end{aligned}
\end{equation}

\noindent where $g_{{j}}$ and $\mathbf{P}$ are the groups of EV users who select the fast-EVCS $e_{j}$ and a Nash stable partition has been found in Section \ref{sec:CSSG}, respectively. $d_{(i,{j})}$ is a binary decision variable that indicates whether or not EV user $b_i$ charges at fast-EVCS $e_{j}$. In the following section, we propose a RL algorithm to solve Eq.(\ref{P:pricing}).

\subsection{Reinforcement Learning Approach}
\label{sec:RL}
\begin{figure}[!ht]
    \centering
    \includegraphics[width=0.99\textwidth]{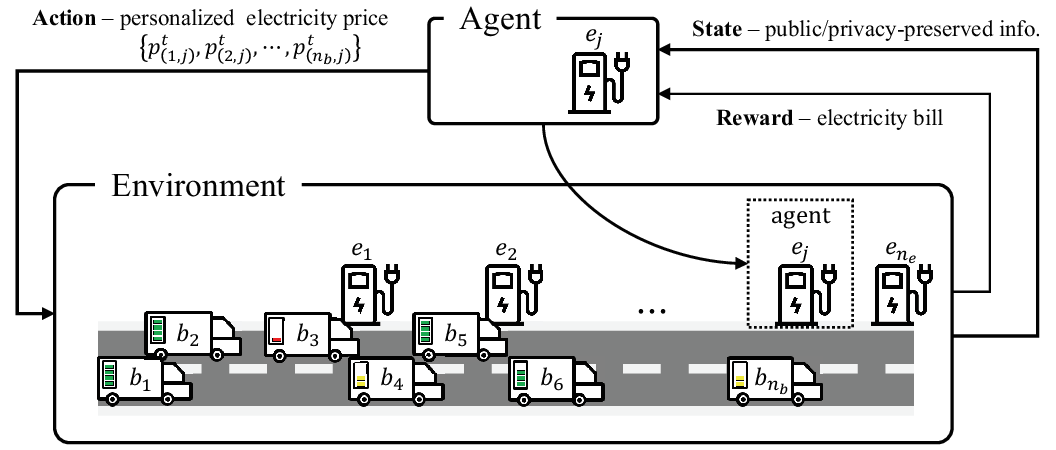}
    \hfil
    \caption{The reinforcement learning framework of PeDP: a fast-EVCS offers personalized electricity prices (action) in the environment, which are interpreted into electricity bills (reward) and the public/privacy-information (state), which are fed back into the fast-EVCS.}
    \label{fig:RLconcept}
\end{figure}

Charging station operators will be able to access a wide variety of information. Also, the charging environment includes many uncertainties, especially due to the drivers' strategic charging decisions. The Reinforcement Learning (RL) approach is known as one of the suitable approaches in an unknown environment with uncertainties where ordinary dynamic programming is not scalable. Reinforcement Learning (RL) is suitable for addressing complex decision-making problems \cite{YING2020210}, \cite{AHAMED2021227}. We consider the fast-EVCS revenue maximization problem in which an agent (AI fast-EVCS, $e_{\bar{j}}$) interacts with the proposed environment in a sequence of actions, observations, and rewards. As shown in Figure \ref{fig:RLconcept}, the AI fast-EVCS serves as the agent; competing fast-EVCSs and strategically behaving EV users are the environment; adjusting personalized electricity prices of the AI fast-EVCS denotes the action that the AI fast-EVCS offers to each corresponding EV user; the public electricity price and the number of available charging slots of the competing fast-EVCSs, and the privacy-preserved information of the EV users represent the state; and the AI fast-EVCS revenue is a reward.
In this paper, the personalized dynamic pricing problem is modeled as a Markov decision process (MDP) because it is a decision-making problem in an unknown stochastic environment. In this MDP model, an agent selects an action to execute with respect to the current state based on its policy. The agent receives an immediate reward and transitions to the next state, in which the reward depends only on EV users' charging station decisions and their SOC in this study. The objective of reinforcement learning is to find the policy that maximizes the expected discounted return

\begin{equation}
    R = \sum_{t=1}^{\infty} \gamma^tr_t 
\end{equation}
\noindent where $r_t$ is the reward at step $t$, and $\gamma \in [0,1]$ is the discount factor.
The key components for the problem to be modeled includes: discrete time $t$, action (personalized electricity price) $A({p}^t_{(i,\bar{j})})$, state $S(v^t_\mathbf{P},\tilde{e}^t_j,\tilde{e}^t_{\bar{j}})$, and reward (revenue) $R(v^t_{\mathbf{P}},\tilde{e}^t_j,\tilde{e}^t_{\bar{j}})$. Depending on the fast-EVCS' pricing policy, the action can be determined. If the fast-EVCS adopts the PeDP, the action is the personalized electricity price, and if the fast-EVCS adopts the PuDP, the action is the public electricity price. Also, the set of observable states might change depending on fast-EVCS' policy.
\begin{itemize}
    \item $t$ is the discrete time at which personalized dynamic price actions are executed. $\Delta t$ is fixed in this study.
    \item ${p}^t_{(i,\bar{j})}$ is personalized electricity price that the AI fast-EVCS $e_{\bar{j}}$ offers at time $t$ for each corresponding EV user $b_i, \, \forall i \in \mathcal{B}$.
    \item $v^t_\mathbf{P}$ indicates the state of the EV users at time $t$, in which the state is $eSOC_{(i,\bar{j})}$.
    \item $\tilde{e}^t_{\bar{j}}$ and $\tilde{e}^t_j$ indicate the state of the AI fast-EVCS (i.e., $\Delta{c}^t_{\bar{j}}$) and the competing fast-EVCSs (i.e., ${p}^t_{(\cdot,j)}$, $\Delta{c}^t_j, \, \forall e_j \in \mathcal{E}$), respectively.
    \item $r_t(\tilde{e}^t_j,\tilde{e}^t_{\bar{j}},v^t_{\mathbf{P}})$ is the AI fast-EVCS revenue, which specifies the expected reward gained by executing action ${p}^t_{(i,\bar{j})}$ for each corresponding EV user $b_i, \, \forall i \in \mathcal{B}$ at a given state.
\end{itemize}
We apply one of the most intuitive and well-known approaches, Q-learning, because the main focus of this study is on personalized pricing using a RL approach (Q1-1). Q-learning is a type of reinforcement learning to learn the policy (known as the Q-function), which is a prediction of reward associated with each action for all actions in each state. The approach is a model-free algorithm that does not use the transition probability distribution and reward function associated with the MDP. In the proposed environment, the fast-EVCS cannot make predictions about the next state and reward before it takes each action due to the EV users' decisions. The algorithm estimates the expected return of executing an action in a given state. These estimated returns are known as Q-values. Q-values are iteratively learned by updating the current Q-value estimate towards the observed reward plus the maximum Q-value over all actions in the resulting state:
\begin{equation}
\begin{aligned}
    Q(&\tilde{e}^t_j,\tilde{e}^t_{\bar{j}},v^t_\mathbf{P},p^t_{(i,\bar{j})}) = Q(\tilde{e}^t_j,\tilde{e}^t_{\bar{j}},v^t_\mathbf{P},p^t_{(i,\bar{j})}) \\
    & + \alpha \Big(r_t(\tilde{e}^t_j,\tilde{e}^t_{\bar{j}},v^t_{\mathbf{P}})+\gamma \max_{p^{t+1}_{(i,\bar{j})}} Q(\tilde{e}^{t+1}_j,\tilde{e}^{t+1}_{\bar{j}},v^{t+1}_\mathbf{P},p^{t+1}_{(i,\bar{j})}) - Q(\tilde{e}^t_j,\tilde{e}^t_{\bar{j}},v^t_\mathbf{P},p^t_{(i,\bar{j})}) \Big).
\end{aligned}    
\end{equation}
Due to the exponentially increasing dimension of traditional Q-learning for the problem with a relatively large number of states and actions, we use a fully connected Artificial Neural Network (ANN) function approximator to estimate the Q-function. We expect that the ANN helps to estimate the unknown underlying function, as it is well known that the ANN is a universal approximator in theory \cite{goodfellow2016deep}. The neural network with one nonlinear hidden layer is parameterized by $\theta$ to represent $Q(\tilde{e}_j,\tilde{e}_{\bar{j}},v_\mathbf{P},p_{(i,\bar{j})};\theta)$\footnote{Note that it is not easy to take deeper and wider networks and say it is better. We have attempted many different settings with different depths and widths of ANNs, but neither the convergence nor the accuracy changed significantly. Based on our simulation environment, a single hidden layer with 128 neurons was sufficient.}.

The learning process is presented in Algorithm \ref{algorithm:Q}, First, we initialize the parameters of the Q-network with $\theta$. As the charging station operates 24/7, the algorithm runs until the convergence criterion is met. Note that the convergence time might be influenced by the parameters of the simulation environments (e.g., EVUs rate, the number of fast-EVCS on the highway, etc.). In most of the cases in this study, it takes 15-30 simulation days to converge. In practical situations, fast-EVCS owners who want to apply the proposed approach might need either a simulation environment or some reference data that applies personalized pricing for an initial PeDP. At each time step, the AI fast-EVCS adjusts the personalized electricity price based on the $\epsilon$-greedy strategy and offers the price to the corresponding EVU. Accordingly, the immediate reward and the next observation are obtained from the environment. The transition is stored sequentially. To update the Q-network, we randomly sample a sequence of transitions $\big \langle S_t, A_t, R_t, S_{t+1} \big \rangle$.

\begin{algorithm}[!ht]
\caption{Q-learning for the personalized dynamic pricing problem}\label{algorithm:Q}
\begin{algorithmic}[1]
\Statex \emph{//Initialization}
\State Initialize environment, replay memory $\mathcal{N}$, $t=1$
\State Initialize Q-network with $\theta$
    \While{Stopping criterion has not been met}
    \State Choose an action $A_t$ at state $S_t$ with probability $\epsilon$
    \State otherwise choose action $A_t=\operatorname*{max}_{A} Q^{\pi}(S_t, A; \theta)$
    \State Execute action $A_t$ in the environment Algorithm \ref{algorithm:GRAPE}
    \State Observe reward $R_t$ and $S_{t+1}$
    \State Store transition $\big \langle S_t, A_t, R_t, S_{t+1} \big \rangle$ in $\mathcal{N}$
    \State Sample random mini-batch of transitions
    \State $\big \langle S_l, A_l, R_l, S_{l+1} \big \rangle$ from $\mathcal{N}$
    \State Compute target ${tg}_l = R_l + \gamma \max_{A_l} {Q}(S_{l+1}, A'; \theta)$
    \State Perform the gradient of ${\big ({tg}_l-Q(S_l,A_l; \theta)\big )}^2$ to update $\theta$
    \State $t = t + 1$
    \EndWhile
\end{algorithmic}
\end{algorithm}

\begin{figure}[!ht]
    \centering
    \includegraphics[width=0.8\linewidth]{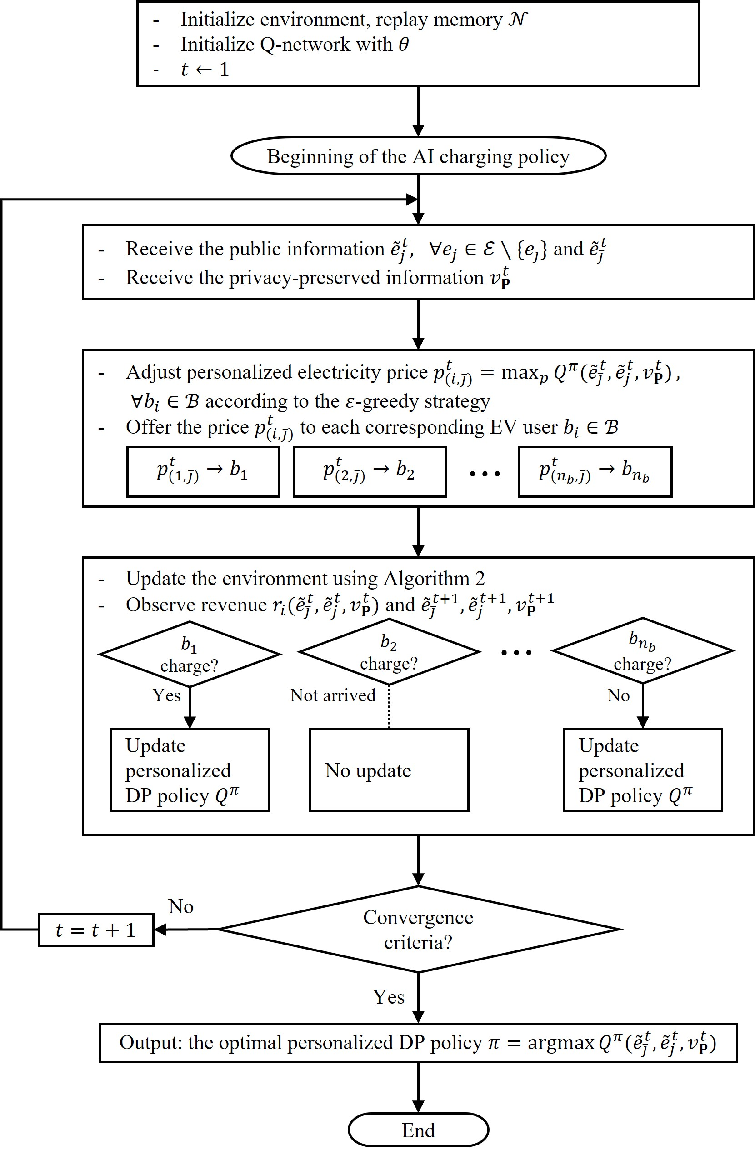}
    \caption{Flowchart for implementing the Q-learning algorithm to the personalized dynamic pricing problem (training stage)}
\label{fig:AlgoFlow}
\end{figure}

The entire process of the proposed approach is presented in Figure \ref{fig:AlgoFlow}. First, we initialize the environment using Algorithm \ref{algorithm:GRAPE}, replay memory, and Q-network weights $\theta$. We adopt the replay memory (experience replay) to have the advantages of improving learning speed and mitigating the undesirable temporal correlations \cite{Mnih2015}. Once the AI fast-EVCS starts to apply the proposed approach, it receives public and privacy-preserved information from competing fast-EVCSs and EV users, respectively. Then, the AI fast-EVCS $e_{\hat{j}}$ adjusts personalized electricity price $p^t_{(i,\bar{j})}$ to corresponding EV user $b_i$ using the $\epsilon-$greedy policy, which randomly chooses an action with probability $\epsilon$ and most of behaves in a greedy manner. We set $\epsilon$ as 0.2, and its discounting factor is 0.99 for all numerical experiments in this study. It offers the price $p^t_{(i,\bar{j})}$ to each corresponding EV user. The environment, which includes EV users' charging station selections and competing fast-EVCSs public electricity prices and the number of available charging slots, is updated using Algorithm \ref{algorithm:GRAPE}. It observes individual revenue from each EV user and updates pricing policies: 1) Yes (arrival and charging) - update pricing policy; 2) Not arrived - no update; and 3) No (pass by the charging station) - update pricing policy. During the training stage, the pricing policy of the AI fast-EVCS is trained only when EV user $b_i$ is charging at the AI fast-EVCS or when it has passed the AI fast-EVCS. To avoid excessive training time, this process runs until the criterion is met, $\Delta L = |L^t(\theta) - L^{t-1}(\theta)| \leq \lambda$ where $\lambda$ (=0.005) is a stopping criterion, in which the criterion is the loss difference between the last two iterations. Finally, the AI fast-EVCS will obtain the PeDP.

\subsubsection{Convergence}
\label{sec:convergence}
The convergence of Algorithm \ref{algorithm:Q} is an important concept to ensure that the learning process produces a pricing policy that achieves the optimal performance in the charging environment. One way to ensure the convergence of the RL approach is known as Greedy in the Limit of Infinite Exploration (GLIE). GLIE is a property of the learning algorithm that guarantees convergence of an optimal policy under certain conditions. It requires that the reward function is stationary and the state-action space is finite, and the agent infinitely explores the environment, in which case we use the stopping criterion $\Delta L$ instead in order to avoid excessive exploration. Also, as the number of iterations of the learning process increases, the probability of a random action decreases. In the limit, the agent selects the greedy action with probability 1, and thus the learning process converges to the optimal policy.

\section{Numerical simulations}
\label{sec:simulation}

For ease of illustration, we conduct the simulations in either the duopoly market (two fast-EVCSs) or the oligopoly market (three fast-EVCSs) with a large number of EV users after training AI fast-EVCS by tuning the hyperparameters of the function approximator. Note that we use the term ``simulation (with training or validation)'' for ``training'' or ``validation'' widely used in machine learning society because we conduct them in the simulation environment only. It is assumed that the specifications of the fast-EVCSs and the EVs in the study are the same as shown in Table \ref{tab:EnvSetting}. For the sake of comparison, the time cycle is divided into 24(hours)$\times$60(minutes). Electricity prices are determined according to the pricing policy of each charging station. Once an AI fast-EVCS has decided to use PuDP or PeDP, they follow that policy and do not change. In the case of the PeDP, the AI fast-EVCS offers personalized electricity prices every minute to corresponding EV users. The prices are valid until the EV user enters the AI fast-EVCS or is no longer interested in the AI fast-EVCS (i.e., they are valid until the EV users pass the AI fast-EVCS). In the case of the PuDP, unlike the PeDP, the AI fast-EVCS broadcasts the public electricity price every minute\footnote{Note that although the time interval for updating prices can be varied depending on pricing policies, operators, etc., the social agreement between customers and business owners should be discussed for the fairness of the service later.}. Conventional fast-EVCS that adopts the gas station's pricing policy broadcasts the public electricity price every half-hour based on Algorithm \ref{algorithm:Bertrand}. Note that the time interval for the price update needs social agreements as it might create confusion and fear of uncertainties when the interval is very short or not defined, and EV users should change their decisions often. However, in this study, we exclude the EV users' confusion and fear.

\begin{table}[!ht]
\caption{Environment settings}
\scriptsize
\label{tab:EnvSetting}
\centering
\begin{tabular}{r|c}
\hline \hline
\multicolumn{2}{c}{\textbf{Environment for fast-EVCSs}}                 \\ \hline
\textbf{\# of fast-EVCSs}                   &\{ $e_1, e_2, \cdots$ \}  \\ \hline
\textbf{Charger type}         & Level 3 (DC charger)      \\ \hline
\textbf{Efficiency} [SOC/min]      & 0.0267                    \\ \hline
\textbf{\# of charging slots}               & 4 for each                \\ \hline
\textbf{Wholesale price} [\euro{}/kWh] & 0.35 \cite{Cazz2018}   \\ \hline
\textbf{Regulated price} [\euro{}/kWh] & 0.55 \cite{Cazz2018}   \\ \hline
\hline

\multicolumn{2}{c}{\textbf{Environment for EVs\footnote{Note that, we utilize the Tesla Semi performance data for the numerical simulations \cite{teslasemi2020}.}}}                   \\ \hline
\textbf{Travel Range} [km]               & 800           \\ \hline
\textbf{Initial SOC}              & Random btw 0.25 and 0.75               \\ \hline
\textbf{Speed of EVs} [km/h]               & 100           \\ \hline
\textbf{Maximum SOC}                    & 0.8           \\ \hline
\textbf{Utility function}                   & Eq.(\ref{monetaryUtility}) or Eq.(\ref{monetaryUtilityNoWait})\\ \hline
\end{tabular}%
\end{table}

It should be pointed out that all the parameters and environmental settings in the numerical simulations are specific and can be changed according to other factors such as the electricity market, the characteristics of the fast-EVCSs and the EV users, etc. Namely, the proposed approach can be suitably applied in other given environment settings. The hyperparameters used for the training in this study are as shown in Table \ref{tab:HyperPara}.

\begin{table}[!ht]
\caption{Hyperparameters for training}
\scriptsize
\label{tab:HyperPara}
\centering
\begin{tabular}{r|c}
\hline
\textbf{Learning rate}                   & 0.01  \\ \hline
\textbf{Epsilon ($\epsilon$)}         & 0.15      \\ \hline
\textbf{Discount factor}      & 0.99                    \\ \hline
\end{tabular}%
\end{table}

Using the optimal pricing policy obtained through the proposed approach in the given environmental setting, 30-day simulations are performed. At each iteration of the simulation, the AI fast-EVCS receives the public/privacy-preserved information; then the AI fast-EVCS adjusts the personalized electricity prices and offers them to corresponding EV users based on the optimal pricing policy; and the AI fast-EVCS receives the rewards. When electricity prices are provided by each fast-EVCS, EV users select a fast-EVCS in a way to maximize the monetary utility function.

The following sections present 30-day simulation results to validate the performance of the proposed PuDP and investigate the importance of information usage from three different perspectives: (Case 1) from the system perspective --- we present the importance of expected waiting time in the EV charging market by comparing the proposed PeDP and the classic Bertrand competition model; (Case 2) from the fast-EVCS perspective --- we verify the performance of the PeDP and compare the simulation results for various information sets to analyze the importance of using each information; (Case 3) from the EV user perspective --- we show a privacy-preserved information misuse case in the EV charging market. Note that, in the numerical simulations, all the environment settings are the same as described in Table \ref{tab:EnvSetting} unless it is mentioned otherwise.

\subsection{Case 1: from the system perspective}
\label{sec:case1}

In the EV charging market, not only the electricity price but also the expected waiting time play a very important role in the EV user's charging station decision. In this numerical simulation, we show the importance of the expected waiting time by comparing the following three cases. First, we show the conventional fast-EVCSs pricing competition using the classic Bertrand competition model in \ref{app:BertrandAlgorithm}. Second, we show the pricing competition in the duopoly market using the proposed PuDP, where a utility function of the EV users is Eq.(\ref{monetaryUtilityNoWait}) (i.e., the EV users only consider the electricity price). 
\begin{equation}
    \label{monetaryUtilityNoWait}
    u^t_{(i,j)} = k_\text{s}\frac{\check{s}_{ij}(c_\text{f} + p^t_{(i,j)}\Delta{SOC_{(i,j)}}\text{B})}{\Delta{SOC_{(i,j)}}}.
\end{equation}

\begin{figure}[!ht]
    \centering
    \includegraphics[width=0.95\linewidth]{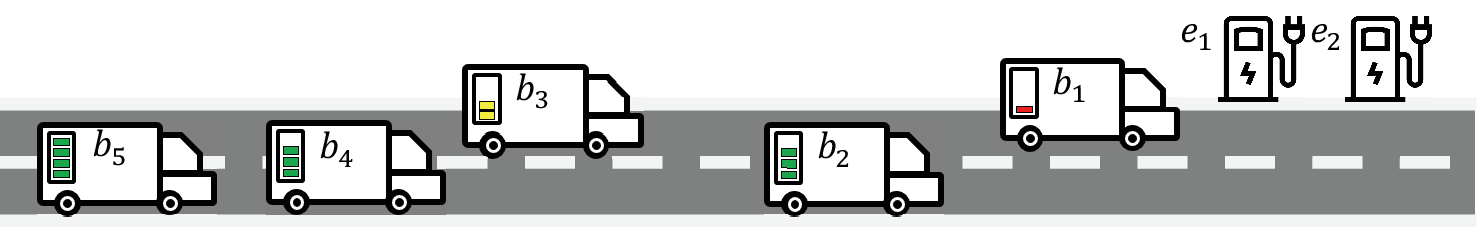}
    \caption{System model for case 1 and case 2. Two fast-EVCSs are located at the same place [$e_1 = 250$km, $e_2 = 250$km]}
    \label{fig:sysModelCase1}
\end{figure}

\noindent We show the simulation results with the expected waiting time using the proposed approach in the same environment. The utility function of the EV users is Eq.(\ref{monetaryUtility}). Figure \ref{fig:sysModelCase1} is the simulation environment where two fast-EVCSs are located at the same place and a large number of EVs are dynamically entering the highway and charging at one of the two fast-EVCSs necessarily. 

\begin{figure}[!ht]
    \centering
    \includegraphics[width=0.5\linewidth]{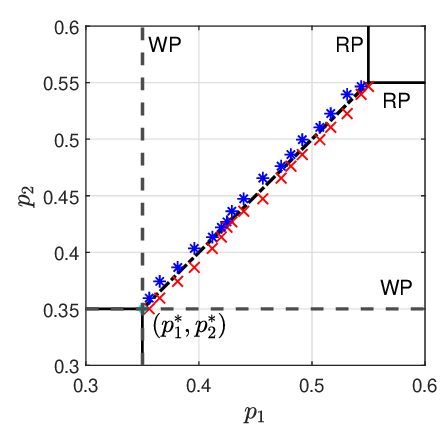}
    \caption{Best response price trajectories for both fast-EVCSs ($e_1=\mlq\times\mrq$ (red), $e_2=\mlq\ast\mrq$ (blue)). RP: Retail Price, and WP: Wholesale Price}
\label{fig:bestRespFunc}
\end{figure}

\begin{figure}[!ht]
    \centering
    \subfloat[]{
      \includegraphics[width=0.3\linewidth]{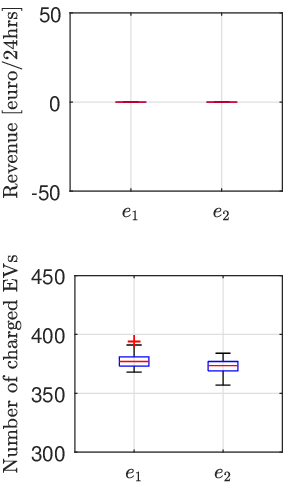}
    }
    \subfloat[]{
      \includegraphics[width=0.3\linewidth]{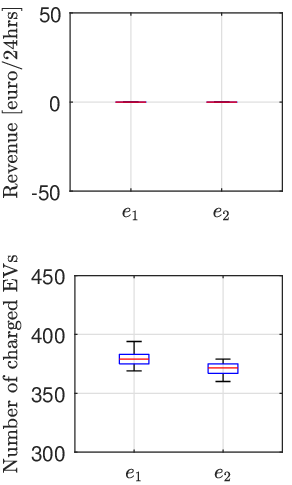}
    }
    \subfloat[]{
      \includegraphics[width=0.3\linewidth]{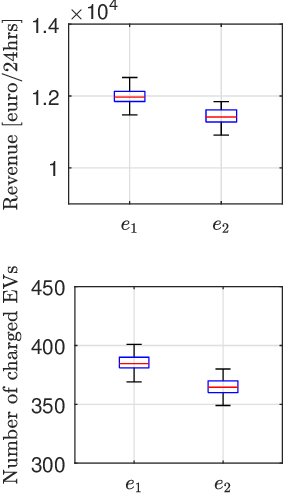}
    }
    \caption{Boxplot of 30-day numerical simulation results for each case: (a) Classic Bertrand competition model; (b) PuDP without expected waiting time; (c) PuDP with expected waiting time.}
\label{fig:boxPlot_case1}
\end{figure}

\noindent \textbf{- Classic Bertrand competition model (no waiting): }
In the classic Bertrand competition model, it is well known that two firms' competitive pricing converges to the marginal cost (wholesale price in this study), which is a Nash equilibrium price. In order to show the price competition, we utilize Algorithm \ref{algorithm:Bertrand} in \ref{app:BertrandAlgorithm}, in which the algorithm is based on the classic Bertrand competition model. Figure \ref{fig:bestRespFunc} shows that when the two fast-EVCSs start to compete at the regulated price, the highest price, the optimal electricity prices of both fast-EVCSs converge to the wholesale price, which is equal to the equilibrium price. Figure \ref{fig:boxPlot_case1}(a) shows the 30-day simulation results when both of the fast-EVCSs set the public electricity prices as the equilibrium price. Since their public electricity prices are equal to the wholesale price, no revenue is obtained even if both fast-EVCSs have EV customers.

\noindent \textbf{- PuDP without expected waiting time:}
In the same environment, we apply the PuDP without expected waiting time to both fast-EVCSs, in which Eq.(\ref{monetaryUtilityNoWait}) is used as the utility function of the EV users. The main difference between Bertrand competition and PuDP without expected waiting time is the customer demand model. For the former one, the whole demand is determined by the price, while the demand is determined by the individual driver’s decision for the later one. It is assumed that both fast-EVCSs are AI, and they compete with each other by adjusting only public electricity prices, from which they can access public information. For example, the AI fast-EVCS $e_1$ can observe $( p^t_{(\cdot, 2)}, \Delta c^t_2 )$ and the AI fast-EVCS $e_2$ can observe $( p^t_{(\cdot, 1)}, \Delta c^t_1 )$. In other words, each fast-EVCS determines the public electricity price based on the information. The pricing policies obtained through the proposed approach converge to the wholesale electricity price for both AI fast-EVCSs. In Figure \ref{fig:boxPlot_case1}(b), therefore, no revenue is obtained for both fast-EVCSs. This is due to the utility function that changes all charging demands with a small price difference, like the Bertrand model.

\begin{figure}[!ht]
    \centering
    \includegraphics[width=0.99\linewidth]{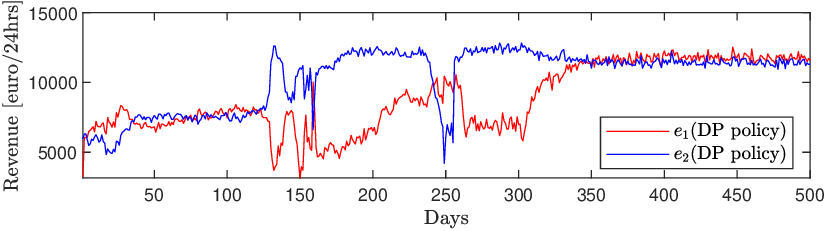}
    \caption{Training trajectories of two AI fast-EVCSs' revenue where the proposed monetary utility function Eq.(\ref{monetaryUtility}) is utilized for the EV users' utility.}
\label{fig:trainTrend_DP}
\end{figure}

\noindent \textbf{- PuDP policy with expected waiting time:} In this case, we also apply the PuDP policy to the same environment as in the two previous cases. In order to show the importance of the expected waiting time to the EV charging market, we use Eq.(\ref{monetaryUtility}) as the utility function of the EV users. As in the previous case, each AI fast-EVCS can observe the number of available charging slots and the public electricity prices. Unlike the two previous pricing policies, when the EV users consider the expected waiting time, the pricing policies of both AI fast-EVCSs converge to the maximum electricity price. The training trajectories of two charging stations are shown in Figure \ref{fig:trainTrend_DP}. From the figure, we can observe that both of the charging stations have been stable since day 350. Before the point, we can see that $e_1$ makes high revenue when $e_2$ makes low revenue, and vice versa. It is mostly because of the electricity price. For example, $e_1$ will make high revenue when the electricity price of $e_1$ is lower than $e_2$'s electricity price (Q1-8).

\noindent The results are due to the expected waiting time, also, the AI fast-EVCSs empirically learn that EV users come to their charging stations for charging in certain situations. For example, even though the public electricity price of the AI fast-EVCS $e_1$ is higher than the personalized electricity price of the AI fast-EVCS $e_2$, the EV users will select $e_1$ if the value of the estimated waiting time at $e_2$ is greater than the electricity price difference between them. In other words, if the expected waiting time of a competitive fast-EVCS is long, the AI fast-EVCS learns empirically that it is possible to increase the public electricity price without a specific model and information. Figure \ref{fig:boxPlot_case1}(c) shows that in this case, both AI fast-EVCSs yield considerably high revenue.


Due to the long charging time of EVs, the EV charging market has been expected to be different from the petrol station market. From the three comparison simulation results, it is found that the expected waiting time plays a very important role in the pricing policy of the fast-EVCSs. Additionally, we find that the PuDP without expected waiting time converges to the same Nash equilibrium price as the classic Bertrand competition model's Nash equilibrium price in the duopoly EV charging market.

\subsection{Case 2: from the fast-EVCS perspective}
\label{sec:case2}

\begin{figure}[!ht]
    \centering
    \includegraphics[width=0.99\linewidth]{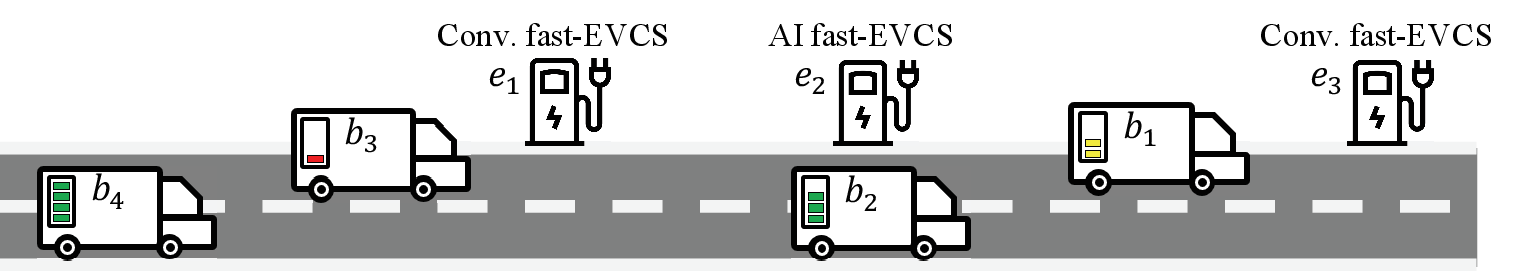}
    \caption{Three fast-EVCSs are located at the different places in the oligopoly market $\lbrack e_1 = 150\text{km}, e_2 = 200\text{km}, e_3 = 400\text{km} \rbrack$.}
\label{fig:case2-1:scheme}
\end{figure}

\noindent \textbf{- Verification of the PeDP:} We show a 30-day simulation result to evaluate the performance of the PeDP in the given environment shown in Figure \ref{fig:case2-1:scheme}. The simulation is conducted based on one AI fast-EVCS ($e_2$[200km]), and two conventional fast-EVCS ($e_1$[150km], $e_3$[400km]). The AI fast-EVCS applies the PeDP while the other two fast-EVCSs adjust the public electricity prices every 30 minutes based on Algorithm \ref{algorithm:Bertrand}\footnote{For exploration in terms of the electricity price, in this simulation, one of two fast-EVCSs ($e_1, e_3$) randomly selects a public electricity price, and another fast-EVCS determines its public electricity price based on only one iteration step of Algorithm \ref{algorithm:Bertrand}.}. The rest of the simulation settings are shown in Table \ref{tab:EnvSetting}.

\begin{figure}[!ht]
    \centering
    \subfloat[]{
      \includegraphics[width=0.22\linewidth]{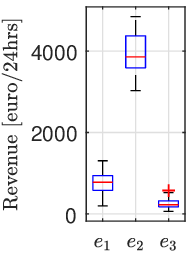}
    }
    \subfloat[]{
      \includegraphics[width=0.22\linewidth]{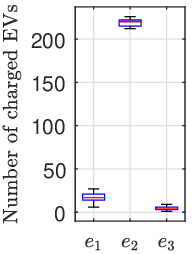}
    }
    \subfloat[]{
      \includegraphics[width=0.22\linewidth]{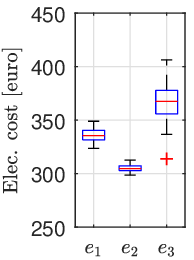}
    }
    \subfloat[]{
      \includegraphics[width=0.22\linewidth]{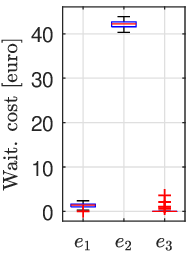}
    }
    \caption{Median values of 30-day simulation results for three fast-EVCSs. (a) accumulative revenue; (b) the number of charged EVs; (c) electricity cost to EV users; (d) waiting cost to EV users.}
\label{fig:3evcsRevenue}
\end{figure}

\begin{figure}[!ht]
    \centering
    \includegraphics[width=0.99\linewidth]{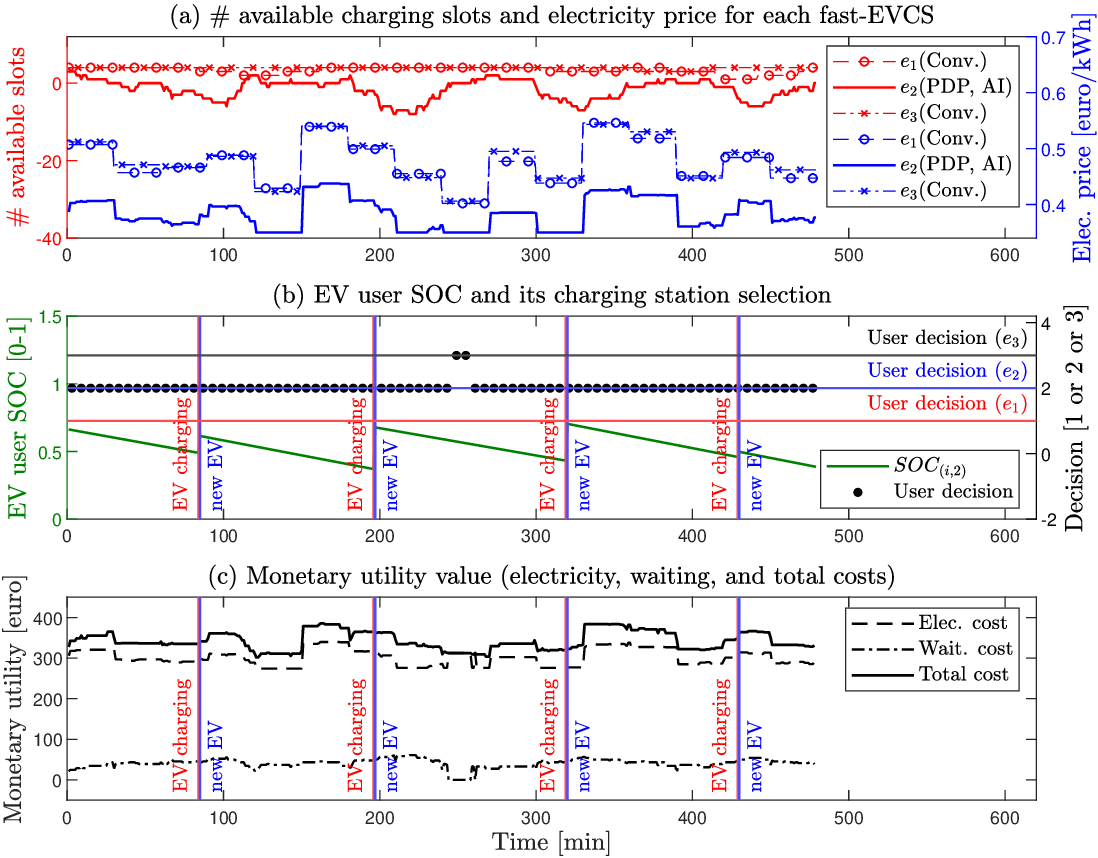}
    \caption{Simulation results of the PeDP for the individual EV users in the given environment shown in Figure \ref{fig:case2-1:scheme}.}
\label{fig:case3IndTraj}
\end{figure}

The main outputs of the simulation are the revenue and the number of charged EV users for each fast-EVCS. Figure \ref{fig:3evcsRevenue}(a) and (b) show the accumulative revenue and the number of charged EVs at each fast-EVCS. Figure \ref{fig:3evcsRevenue}(a) and (b) clearly show that the AI-fast EVCS $e_2$ with the PeDP makes the most revenue among them by attracting most of the EV users. Figure \ref{fig:3evcsRevenue}(c) and (d) show that the EV users who select $e_2$ pay less electricity cost but cost more on waiting time than the EV users who select either $e_1$ or $e_3$. From Figure \ref{fig:case3IndTraj}, we can see the more detailed PeDP. Figure \ref{fig:case3IndTraj}(a) shows the trajectories of the number of available charging slots and the personalized electricity price over time for each fast-EVCS based on their PeDP. Based on the public/privacy-preserved information, such as the number of available charging slots, the public electricity prices of competitors, etc., the AI fast-EVCS adjusts the personalized electricity price and offers it to the corresponding EV user. Each EV user selects one fast-EVCS among them to charge based on the public or personalized electricity prices and the number of available charging slots of the fast-EVCSs, as shown in Figure \ref{fig:case3IndTraj}(b). In other words, the EV users choose fast-EVCSs based on the proposed monetary utility function, Eq.(\ref{monetaryUtility}). Figure \ref{fig:case3IndTraj}(c) shows the monetary utility of the EV users for the fast-EVCS $e_2$. In Figure \ref{fig:case3IndTraj}(b), for example, let us see the decision-making process of the EV user, $b_i$, who enters the highway at 197 minutes with SOC level 0.67 and visits the AI fast-EVCS, $e_2$, at 319 minutes with SOC level 0.43. As shown in Figure \ref{fig:case3IndTraj}(a) the lower personalized electricity price of $e_2$ attracts the EV user despite the lack of available charging slots at the fast-EVCS $e_2$. Exceptionally, even with the lower personalized electricity price and no waiting cost between 244 minutes and 258 minutes, the EV user selects $e_3$, which is 200km further away from $e_2$. This is due to the lack of available charging slots at $e_2$. We can see a small decrease in the total monetary utility value in Figure \ref{fig:case3IndTraj}(c). In other words, EV users always try to make economical choices, and the AI-fast EVCS offers marginal prices to attract EV users.

\noindent \textbf{- Information importance analysis:} From the above simulation results of the performance verification, the AI fast-EVCS adapts to the given environment and finds the optimal pricing policy based on the given information. In this case, we analyze how the public/privacy-preserved information affects pricing policies for maximizing revenue through numerical simulations with different state sets. For the fair and simple comparison as shown in Figure \ref{fig:sysModelCase1}, we place one AI fast-EVCS $e_1$ and one conventional fast-EVCS $e_2$ at the same location, which is 250km from the entrance of the highway. In addition, we fix the public electricity price of the competing fast-EVCS $(p_2=0.52)$ to observe the influence and importance of the different state sets. We find the PeDP or PuDP and conduct 30-day simulations for four cases: Case 2.1: $ s(p^t_{(\cdot,j)})$; Case 2.2: $ s(p^t_{(\cdot,j)}, v^t_{\mathbf{P}}) $; Case 2.3: $ s(p^t_{(\cdot,j)}, \Delta c_j^t, \Delta c_{\bar{j}}^t) $; and Case 2.4: $ s(p^t_{(\cdot,j)}, v^t_{\mathbf{P}}, \Delta c_j^t, \Delta c_{\hat{j}}^t) $, $\forall e_j \in \mathcal{E} \setminus e_{\bar{j}}$. The simulation results are shown in Table \ref{tab:case2-2}. Note that Case 2.1 and 2.3 do not have privacy-preserved information about EV users, so they broadcast the public electricity price to all EV users. That is, the two cases are the PuDP. 

\begin{table*}[!ht]
\centering
\scriptsize
\caption{Median values of 30-day numerical simulation results for the four cases in Figure \ref{fig:sysModelCase1}. Case 2.1: $ s(p^t_{(\cdot,j)})$; Case 2.2: $ s(p^t_{(\cdot,j)}, v^t_{\mathbf{P}}) $; Case 2.3: $ s(p^t_{(\cdot,j)}, \Delta c_j^t, \Delta c_{\bar{j}}^t) $; and Case 2.4: $ s(p^t_{(\cdot,j)}, v^t_{\mathbf{P}}, \Delta c_j^t, \Delta c_{\hat{j}}^t) $}
\begin{tabular}{c|c|c|c|c|c}
\hline
           & \multicolumn{2}{c|}{Fast-EVCSs ($e_1$/$e_2$)} & \multicolumn{3}{c}{EV users' utility to fast EVCSs ($e_1$/$e_2$) {[}euro{]}} \\ \hline
Case       & Revenue {[}euro{]}           & \# of EVs           & Total utility     & Electricity utility     & Waiting utility     \\ \hline
2.1   & (4392.2/0.0) & (193/0) & (359.3/0.0) & (326.1/0.0) & (32.9/0.0) \\
2.2   & (4958.6/0.0) & (193/0) & (373.6/0.0) & (333.4/0.0) & (40.0/0.0) \\
2.3   & (6725.1/447.1) & (186/8) & (385.6/400.93) & (356.7/400.9) & (28.7/0.0) \\
2.4   & (6832.5/1175.2) & (173/20) & (385.4/402.2) & (364.4/401.6) & (20.9/0.3) \\ \hline
\end{tabular}\label{tab:case2-2}
\end{table*}

Case 2.1 shows the lowest revenue while the number of charged EVs is high, as shown in Table \ref{tab:case2-2}. Because of the limited information, the AI fast-EVCS $e_1$ conservatively adjusts the public electricity price. For example, the AI fast-EVCS $e_1$ somehow assumes that the number of available charging slots $\Delta c_2^t$ is always the maximum value because the AI fast-EVCS $e_1$ does not observe $\Delta c_2^t$ for the pricing policy.

In Case 2.2, the AI fast-EVCS $e_1$ receives the privacy-preserved information $v^t_{\mathbf{P}}$, which is $eSOC_{(i,1)}$, $\forall b_i \in \mathcal{B}$. Based on the information, the AI fast-EVCS adjusts the personalized electricity price for each corresponding EV user. This case shows the advantage of observing the privacy-preserved information. The AI fast-EVCS can implicitly consider the fixed cost $c_\text{f}$ in Eq.(\ref{monetaryUtility}) because the cost depends on the $\Delta SOC_{(i,j)}$. For example, the monetary utility of EV user $b_{i}$ is higher than that of EV user $b_{i'}$ when $eSOC_{(i,1)} > eSOC_{({i'},1)}$. Therefore, based on the information, the AI fast-EVCS determines personalized electricity prices to attract customers with 12.9\% more revenue than Case 2.1. 

Case 2.3 shows the simulation results when the AI fast-EVCS $e_1$ considers the number of available charging slots $\Delta c_2^t$ and the public electricity price $p^t_{(\cdot,2)}$ as the state. $\Delta c_2^t$ is directly related to the expected waiting time at $e_2$ which affects the EV user's decision-making. In other words, if the public electricity price gap between two fast-EVCSs, $p^t_{(\cdot,2)} - p^t_{(\cdot,1)}$, is bigger than the waiting cost at $e_1$, which is the latter term of Eq.(\ref{monetaryUtility}), the EV user selects the AI fast-EVCS $e_1$ for charging. Table \ref{tab:case2-2} shows that Case 2.3 yields 53.12\% and 35.63\% higher returns than Case 2.1 and Case 2.2, respectively. In other words, to increase revenue, it is shown that it is more important to utilize the number of charging slots $\Delta c_2^t$ rather than utilizing other information.

In Case 2.4, finally, the AI fast-EVCS $e_1$ considers the public electricity price $p^t_{(\cdot,2)}$ and the number of available charging slots $\Delta c_2^t$ and estimated SOC $eSOC_{(i,1)}, \forall b_i \in \mathcal{B}$ as the state. The AI fast-EVCS has the highest revenue among the other cases, even if the AI fast-EVCS loses some EV users. Nevertheless, the reason for the high revenue can be found in the electricity utility in Table \ref{tab:case2-2}. In Case 2.4, the AI fast-EVCS benefits from EV users by offering higher personalized electricity prices than in any other case. Note that the purpose of the simulation results is not to show a quantitative comparison but to analyze the importance of the information as the states of the AI fast-EVCS. 


\subsection{Case 3: from the EV user perspective}
\label{sec:case3}

\begin{figure}[!ht]
    \centering
    \includegraphics[width=0.99\linewidth]{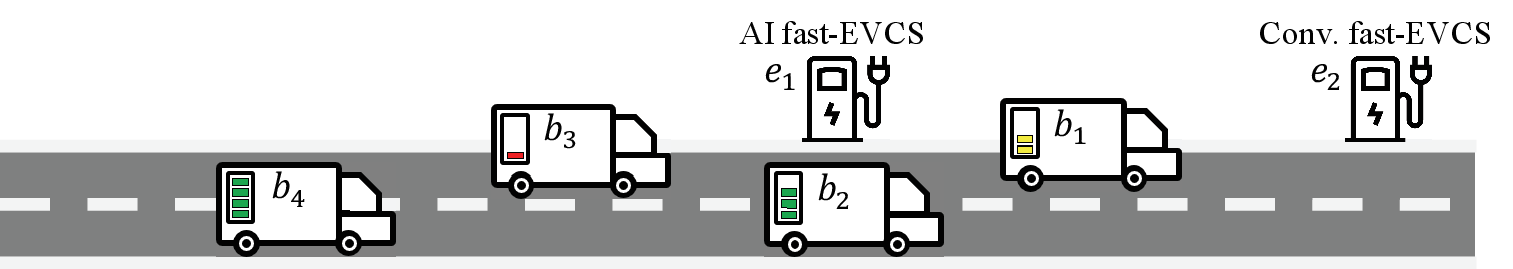}
    \caption{Two fast-EVCSs are located at the different places in the duopoly market $\lbrack e_1 = 200\text{km}, e_2 = 400\text{km} \rbrack$.}
\label{fig:sysModelCase3}
\end{figure}

In the previous two cases, we showed the importance of the information and the performance of the proposed approach. In Case 3, we present a case where even the privacy-preserved information could be abused by the AI fast-EVCS. In the duopoly EV charging market, when two fast-EVCSs are in different locations, the AI fast-EVCS close to the entrance of the highway is likely to misuse the privacy-preserved information for its revenue improvement. We show a misuse case in the duopoly EV charging market using the PeDP in the environment, as shown in Figure \ref{fig:sysModelCase3}. Except for the fast-EVCS locations ($e_1=$200km and $e_2=$400km), the environment is the same as the environment for Case 2.2.

\begin{figure}[!ht]
    \centering
    \subfloat[]{
      \includegraphics[width=0.5\linewidth]{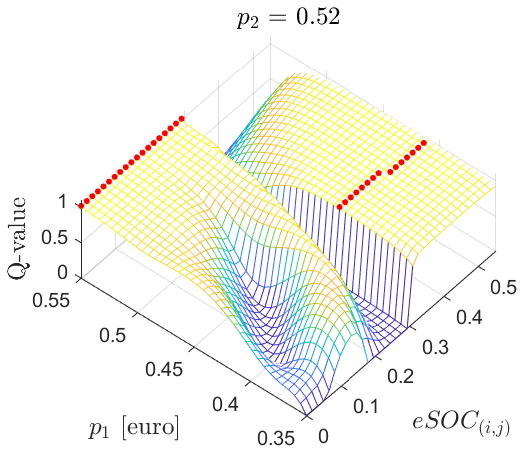}
    }
    \subfloat[]{
      \includegraphics[width=0.5\linewidth]{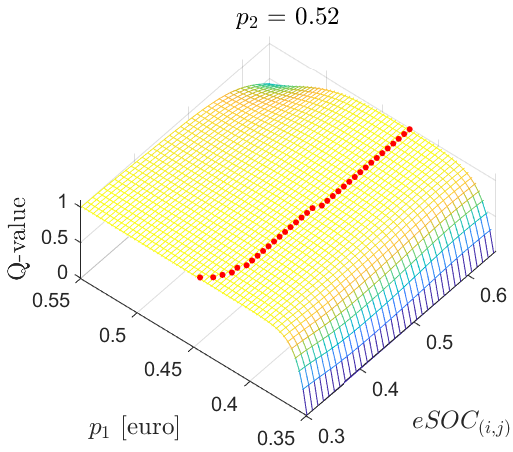}
    }
\caption{The PeDP with regards to the eSOC for (a) the environment shown in Figure \ref{fig:sysModelCase3}; and for (b) the environment for Case 2.2 shown in Figure \ref{fig:sysModelCase1}.}
\label{fig:personalizedDP}
\end{figure}

As similar to Case 2.2, the AI fast-EVCS attracts most of the EV users with marginal peronalized electricity prices. However, due to the low SOC, some EV users cannot reach $e_2$. In other words, these EV users are forced to charge at $e_1$ regardless of the electricity price. In such a situation where two fast-EVCSs are separated from each other, simply sharing eSOC, which is privacy-preserved information, can lead to an unfair pricing policy (i.e., causing a transformation from the duopoly market to the monopoly market). The misused PeDP obtained using the proposed approach for the environment is shown in Figure \ref{fig:personalizedDP}(a). As we discussed, the reason for the dramatically increasing personalized electricity price is obvious. Since the EV users do not have enough SOC to reach $e_2$, they only have the option to charge at $e_1$. The AI fast-EVCS empirically knows the facts and utilizes them. On the other hand, Figure \ref{fig:personalizedDP}(b) shows the PeDP of Case 2.2, which has no abuse of the privacy-preserved information. Therefore, EV users are always in the duopoly market when two fast-EVCSs are located in the same place. From the simulation results, we conclude that sharing the privacy-preserved information, which is the eSOCs of anonymous EVs, could implicitly change the duopoly market to the monopoly market and could be misused by the AI fast-EVCS. Therefore, in order to avoid misuse cases, policymakers should be more careful in permitting the utilization of privacy-preserved information or in the placement of fast-EVCSs such as Figure \ref{fig:case2-1:scheme} (i.e., the oligopoly market).

\section{Discussion, conclusions, and future research}
\label{sec:conclusion}

In this paper, the PeDP algorithm is proposed, wherein the AI fast-EVCS can adaptively adjust the personalized electricity price in the proposed environment using the RL methodology according to the public information from competing fast-EVCSs and the privacy-preserved information from the EV users.

Through the use of the proposed approach, the AI fast-EVCS does not require any specific model of the environment on which an action for personalized electricity price needs to be selected. Instead, the proposed approach utilizes the relationship between states, actions, and rewards for learning through the dynamic interaction between the AI fast-EVCS, the competing fast-EVCSs, and EV users. Therefore, such a learning approach might be useful for managing the fast-charging stations in the near future, where we should be concerned about the electricity shortage. Apart from the methodological findings for the application, the numerical simulation results show 1) the radical difference between conventional refueling and fast-EVCS charging due to the long charging time as shown in Section \ref{sec:case1}; 2) the performance of the PeDP is evaluated and the importance of information is analyzed with the different sets of states as shown in Section \ref{sec:case2}; and 3) the AI fast-EVCSs could misuse the privacy-preserved information of anonymous EVs for maximizing their revenue, which is forbidden in the competitive market as shown in Section \ref{sec:case3}. For the implementation of the proposed approach, users might be required to provide a data set without applying the approach for training for some period of time. As Q-learning is an off-policy RL approach, it is easy to implement, update, or change the policy with a trained model. The computational time to real-time run it is relatively of low cost.

To apply this research to real-world systems, there are still several open questions that need consideration. Obtaining charging station decision-making data from EV users is crucial. In this study and many previous ones, the lack of such information led us to model them and use assumptions. If the decision-making data in real environments were available, it would enable the development of a more meaningful decision model for EV users. Despite the assumptions made in our research and others, the consideration of a 1-D scenario, even on highways, has inherent limitations in many real-world cases. In complex network situations, such as highways, it becomes necessary to address these limitations. Game-based methodologies may face increased modeling complexity in such scenarios, making it challenging to model and find equilibrium. In the case of the methodology we proposed, careful consideration is needed regarding computational complexity. Additionally, our study focused solely on decision-making aspects and did not consider power grid stability. This constraint should be acknowledged and addressed.

In the future, the presented framework could be utilized to investigate the following practical problem: 1) revenue maximization problem in competitive environments with multiple AI fast-EVCSs; 2) electricity pricing policy optimization around the city center by considering the traffic networks; and 3) stable energy distribution problem by pricing. We also suggest readers investigate the following directions: policy gradient-based RL approaches for the personalized dynamic pricing policy; include the following question in the problem statement: ``Which type of pricing policy (PeDP or PuDP) an AI fast-EVCS should select?''; when a service provider owns and operates multiple charging stations, which might require synchronization and/or coordination algorithms; when smart EV users select their most cost-effective EVCS. Finally, such personalized policies should be discussed and agreed upon from a social-ethical equity perspective and approved by authorities.

\section*{Acknowledgment}
\label{sec:Ack}
B.  Kulcsár  acknowledges the  contribution of Transport Area of Advance  at Chalmers  University  of Technology. The  project  has  been  partially  supported by European Comision(F-ENUAC-2022-0003) and Energimyndigheten (2023-00021) through the project E-LaaS:Energy optimal urban Logistics As A Service. 

\appendix
\section{Appendix A: Algorithm for CSSG} 
\label{App:algorithm}
In this appendix, we describe the algorithm for solving the charging station selection game described in Section \ref{sec:systemModel}. \emph{Nash stable partition} is formally defined as follows.

\begin{definition}[Deviated partition]
Given an instance of CSSG defined as a tuple $(\mathcal{B},\mathcal{E},\mathcal{P})$, EV user $b_i$ may unilaterally deviate to group $g_j$ since the user prefers to fast-EVCS $e_j$ in the context. We refer to the resultant partition as a \emph{deviated partition}, denoted by $\mathbf{P}_{i \rightarrow j}$.
\end{definition}

\begin{definition}[Nash stable partition]
A partition $\mathbf{P}$ is $\emph{Nash stable}$ if, for every EV user $b_i \in \mathcal{B}$, it holds that $u_i(\mathbf{P}) \ge u_i(\mathbf{P}_{i \rightarrow j}), \; \forall e_j \in \mathcal{E}$.
\end{definition}

Given an instance of charging station selection game (CSSG), the algorithm to find a Nash stable partition is presented in Algorithm \ref{algorithm:GRAPE}. Here, superscripts indicate local information known to EV users. 

\begin{algorithm}[!ht]
\caption{Decision-making of each EV user $b_i$}\label{algorithm:GRAPE}
\begin{algorithmic}[1]
\Function{CSSG}{$\mathbf{P}$}
\Statex \emph{// Initialization}
\State $\mathbf{P}^i \leftarrow \mathbf{P}$; $r^i \leftarrow 0$; $s^i \leftarrow 0$; $\mathsf{satisfied}^i \leftarrow \mathbf{0}^{n_\alpha \times 1}$ \label{line:local_variable_1}
    \Repeat

 	\Statex \emph{// Make a new decision if necessary}
	\If{$\mathsf{satisfied}^i[i] = 0$} \label{line:decision_making}

	\State $\hat{\mathbf{P}}^i \leftarrow \arg\max_{\forall \mathbf{P}^i_{i \rightarrow j} \in \widetilde{\mathbf{P}}^i_i} u_i(\mathbf{P}^i_{i \rightarrow j} )$ \label{line:choose_best}
	\If{$\hat{\mathbf{P}}^i \neq \mathbf{P}^i$} \label{line:decision_1}
	\State $\mathbf{P}^i \leftarrow \hat{\mathbf{P}}^i$
	\State $r^i \leftarrow r^i + 1$; $s^i \in \mathrm{unif}[0,1]$
	\State $\mathsf{satisfied}^i \leftarrow \mathbf{0}^{n_\alpha \times 1}$
	\label{line:iteration_increase}
	\EndIf \label{line:decision_2}
	\State $\mathsf{satisfied}^i[i] \leftarrow 1$ \label{line:satisfied}
	\EndIf \label{line:decision_making2}
	
 	\Statex \emph{// Distributed Mutual Exclusion}
	\State Broadcast $M^i = \{r^i, s^i, \mathbf{P}^i, \mathsf{satisfied}^i\}$ 
	\State to the selected EVCS and
	\State receive $M^k$ from EVCSs $\forall e_j \in \mathcal{E}$ \label{line:communication}

	\State Collect all the messages $\mathcal{M}^i_{rcv}=\{M^i, \forall M^k\}$ \label{alg_mutex:line0}
	
	\For{each message $M^k \in \mathcal{M}^i_{rcv}$} \label{alg_mutex:line1}
	\If{($r^k > r^i$) or ($r^k = r^i$ \& $s^k > s^i$)} 
		\State $M^i \leftarrow M^k$
	\EndIf 
	\EndFor \label{alg_mutex:line2}	

	\Until{$\mathsf{satisfied}^i = \mathbf{1}^{n_\alpha \times 1}$} 
\State \Return $\mathbf{P}^i$	
\EndFunction
\end{algorithmic}
\end{algorithm}

Every EV user $b_i \in \mathcal{B}$ has local variables for its decision such as $\mathbf{P}^{i}$, $r^i$, $s^i$, and $\mathsf{satisfied}^i$ (line \ref{line:local_variable_1}). Here, $\mathbf{P}^{i}$ is the EV user's locally known partition; $r^i \in \mathbb{Z}^{>}$ is a positive integer variable to represent how many times $\mathbf{P}^{i}$ has evolved; and $s^{i}\in [0,1]$ is a uniform random variable that is generated whenever $\mathbf{P}^{i}$ is updated; and $\mathsf{satisfied}^i$ is a binary array that includes the information whether or not each EV user is satisfied with $\mathbf{P}^{i}$ such that it does not want to deviate from its current fast-EVCS. Given $\mathbf{P}^{i}$, EV user $b_i$ examines which fast-EVCS is the most preferred among other fast-EVCSs, assuming that the other EV users stay at the current fast-EVCSs (line \ref{line:choose_best}). Then the EV user selects the newly found fast-EVCS if it is more preferred than the current fast-EVCS. In this case, the EV user updates $\mathbf{P}^{i}$ to reflect its new decision, increases $r^i$, and generates a random time stamp $s^i$ (line \ref{line:decision_1}-\ref{line:decision_2}). In any case, the EV user ascertained that the current fast-EVCS is the most preferred one, so the EV user satisfies with $\mathbf{P}^{i}$ (line \ref{line:satisfied}). Then, EV user $b_i$ sends its status to the selected fast-EVCS via a message $M^i:=\{ r^i,s^i,\mathbf{P}^i,\mathsf{satisfied}^i \}$, and receives $M^j$ from all fast-EVCSs (line \ref{line:communication}).

Since every EV user updates their locally known partition obtained from the fast-EVCSs simultaneously, one of the partitions should be considered as if it were the partition updated by a deciding EV user at the previous iteration (refer to this partition as \textit{the valid partition}). The distributed mutual exclusion part of Algorithm \ref{algorithm:GRAPE} enables the EV users to identify the valid partition. Once each EV user $b_i$ collects all the messages from all fast-EVCSs $\mathcal{M}^i_{rvc}$ (line \ref{alg_mutex:line0}). Based on the message set, the EV user identifies whether or not their partition $\mathbf{P}^i$ is valid. In the case if there are partitions $\mathbf{P}^k$ such that $r^i < r^k$, then the EV user judges that partition $\mathbf{P}^k$ is more valid then $\mathbf{P}^i$. When $r^i = r^k$, a higher time stamp $s^i$ and $s^k$ determine the valid partition. Now, since $\mathbf{P}^k$ is judged as more valid, EV user $b_i$ needs to re-examine whether or not there is a more preferred fast-EVCS in the next iteration (line \ref{alg_mutex:line1}-\ref{alg_mutex:line2}). Depending on $\mathsf{satisfied}^i$, each EV proceeds the decision-making process again until $\mathsf{satisfied}^i = \mathbf{1}^{n_\alpha \times 1}$.

In Algorithm \ref{algorithm:GRAPE}, the behavior of each EV user is to pursue its highest monetary utility greedily and unilaterally. EV users executing this straightforward local algorithm can find a stable outcome as a collective behavior under certain conditions. One of them is called \emph{SPAO (Single-Peaked-At-One)}\cite{Jang2018g}, and this condition enables the agents in Algorithm \ref{algorithm:GRAPE} to converge to a Nash stable partition. In the context of this paper, SPAO can be rephrased as follows:

\begin{definition}[SPAO condition]\label{condition:SPAO}
Given $\mathbf{P}$, for EV user $b_i$, let $\widetilde{\mathbf{P}}_{-i} = \cup_{\forall b_l \in \mathcal{B} \setminus \{b_i\}} \widetilde{\mathbf{P}}_{l}$ denote all the possible deviated partitions by any of all the other EV users.
The individual utility of EV user $b_i$ is \emph{SPAO (Single-Peaked-At-One)}
if it holds for any $\mathbf{P}$ that $u_i(\mathbf{P}) \ge u_i(\mathbf{P}')$ for $\forall \mathbf{P}' \in \widetilde{\mathbf{P}}_{-i}$ such that ${|g_{\mathbf{P}(i)}| < |g_{\mathbf{P}'(i)} }|$.  
\end{definition}

In other words, the condition can be interpreted as that each EV user utility is monotonically decreased whenever a new EV user joins the same group where the former belonged to, regardless of the current partition.
This fact implies that EV users satisfying this condition tend to avoid crowdedness.

\begin{lemma}\label{lemma:basic}
If an instance $(\mathcal{B},\mathcal{E},\mathcal{P})$ in which the current partition is Nash stable holds SPAO, the new instance $(\hat{\mathcal{B}},\mathcal{E},\hat{\mathcal{P}})$ that has a new EV $b_q \notin \mathcal{B}$ holding SPAO with regard to every EVCS $e_j \in \mathcal{E}$ also, (1) satisfies SPAO; (2) contains a Nash stable partition; and (3) the maximum number of iterations required to converge to a Nash stable partition is $|\hat{\mathcal{B}}|$.
\end{lemma}

\begin{theorem}[Existence of Algorithm \ref{algorithm:GRAPE}]\label{Theorem:existence}
If an instance $(\mathcal{B},\mathcal{E},\mathcal{P})$ holds SPAO condition, then a \textit{Nash stable} partition always exists. 
\end{theorem}

\begin{theorem}[Convergence of Algorithm \ref{algorithm:GRAPE}]\label{Theorem:convergence}
If an instance $(\mathcal{B},\mathcal{E},\mathcal{P})$ holds SPAO condition, then the number of iterations required to determine a \textit{Nash stable} partition is at most $|\mathcal{B}| \cdot(|\mathcal{B}|+1)/2$. 
\end{theorem}

More details of the proofs of Lemma \ref{lemma:basic}, Theorem \ref{Theorem:existence} and Theorem \ref{Theorem:convergence} can be found in \cite{Jang2018g, Bae2020}

\section*{Appendix B: Algorithm for the classic Bertrand competition model}
\label{app:BertrandAlgorithm}

This model is based on the following assumptions: 1) at least two fast-EVCSs providing the same quality electricity (i.e., homogeneous goods); 2) they cannot cooperative each other; 3) they competes by adjusting public electricity prices simultaneously; 4) EV users want to charge from a fast-EVCS with a lower public electricity price; and 5) if the fast-EVCSs set the same public electricity price, EV users' demand is splits evenly between them. We suggest a simple algorithm to find the equilibrium price of the model \cite{yadati2009}.

\begin{algorithm}[!ht]
\caption{Algorithm for the Bertrand model}\label{algorithm:Bertrand}
\begin{algorithmic}[1]
\State First day $p_1$ and $p_2$ are set RP
    \For{day $=2, \text{end day}$}
    \If{$p_1 >= p_2$}
    \State $ p_1 = p_2 - \epsilon $
    \ElsIf{$p_1 < p_2$}
    \State $ p_2 = p_1 - \epsilon $
    \EndIf
    \If{$p_i <$ WP}
    \State $ p_i =$ WP
    \EndIf
    \EndFor
\end{algorithmic}
\end{algorithm}
\noindent where we assume that $\epsilon$ is very small value between. In the first day, both of the fast-EVCS set public electricity prices to the regulated price. Any of two decrease its public electricity price if the price is higher than that of another fast-EVCS. The public electricity prices of the two fast-EVCSs converge to the marginal electricity price, which is a Nash equilibrium price.

\section*{Appendix C: Loss function minimization}
\label{app:lossFunction}

The artificial neural network is trained by minimizing the following loss function:
\begin{equation*}
    L(\theta) = {\big(y^{tg}_k-Q(\tilde{e}^t_j,\tilde{e}^t_{\bar{j}},v^t_\mathbf{P},p^t_{(i,\bar{j})};\theta)\big)}^2
\end{equation*}
\noindent where $y^{tg}_t=r_i(\tilde{e}^t_j,\tilde{e}^t_{\bar{j}},v^t_{\mathbf{P}})+\gamma \max_{p^{t+1}_{(i,\bar{j})}} Q(\tilde{e}^{t+1}_j,\tilde{e}^{t+1}_{\bar{j}},v^{t+1}_\mathbf{P},p^{t+1}_{(i,\bar{j})};\theta^-)$ denotes the target value of action $p^{t}_{(i,\bar{j})}$ given state $\tilde{e}^t_j,\tilde{e}^t_{\bar{j}},v^t_{\mathbf{P}}$. Here, $\theta^-$ is cloned from $\theta$ every fixed number of iterations. At each training iteration $t$, an experience $n_t=\big \langle \tilde{e}^t_j,\tilde{e}^t_{\bar{j}},v^t_\mathbf{P},p^t_{(i,\bar{j})},r_t(\tilde{e}^t_j,\tilde{e}^t_{\bar{j}},v^t_{\mathbf{P}}),\tilde{e}^{t+1}_j,\tilde{e}^{t+1}_{\bar{j}},v^{t+1}_\mathbf{P}\big \rangle$ is sampled uniformly from the replay memory $\mathcal{N}$ \cite{Mnih2015}. When the AI fast-EVCS selects and offers personalized electricity prices according to an $\epsilon-$greedy. Note that this algorithm solves the problem directly using samples from the proposed environment without explicitly modeling the environment (\textit{model-free}).

\bibliography{library}

\begin{thebibliography}{10}
\expandafter\ifx\csname url\endcsname\relax
  \def\url#1{\texttt{#1}}\fi
\expandafter\ifx\csname urlprefix\endcsname\relax\def\urlprefix{URL }\fi
\expandafter\ifx\csname href\endcsname\relax
  \def\href#1#2{#2} \def\path#1{#1}\fi

\bibitem{Cazz2018}
P.~Cazzola, M.~Gorner, S.~Scheffer, R.~Scuitmaker, J.~Tattini, {Nordic EV Outlook 2018}, Tech. rep., International Energy Agency (2018).

\bibitem{pereira2022short}
M.~Pereira, A.~Lang, B.~Kulcs{\'a}r, Short-term traffic prediction using physics-aware neural networks, Transportation research part C: emerging technologies 142 (2022) 103772.

\bibitem{DABIRI2015585}
A.~Dabiri, B.~Kulcsár, Freeway traffic incident reconstruction – a bi-parameter approach, Transportation Research Part C: Emerging Technologies 58 (2015) 585--597.

\bibitem{7539538}
A.~Dabiri, B.~Kulcsár, H.~Köroğlu, Distributed lpv state-feedback control under control input saturation, IEEE Transactions on Automatic Control 62~(5) (2017) 2450--2456.

\bibitem{CSIKOS2017429}
A.~Csikós, B.~Kulcsár, Variable speed limit design based on mode dependent cell transmission model, Transportation Research Part C: Emerging Technologies 85 (2017) 429--450.

\bibitem{CSIKOS2017120}
A.~Csikós, T.~Charalambous, H.~Farhadi, B.~Kulcsár, H.~Wymeersch, Network traffic flow optimization under performance constraints, Transportation Research Part C: Emerging Technologies 83 (2017) 120--133.

\bibitem{pereira2022parameter}
M.~Pereira, P.~B. Baykas, B.~Kulcs{\'a}r, A.~Lang, Parameter and density estimation from real-world traffic data: A kinetic compartmental approach, Transportation Research Part B: Methodological 155 (2022) 210--239.

\bibitem{Chen2016}
Z.~Chen, F.~He, Y.~Yin, \href{https://linkinghub.elsevier.com/retrieve/pii/S0191261516303319}{{Optimal deployment of charging lanes for electric vehicles in transportation networks}}, Transportation Research Part B: Methodological 91 (2016) 344--365.
\newblock \href {https://doi.org/10.1016/j.trb.2016.05.018} {\path{doi:10.1016/j.trb.2016.05.018}}.
\newline\urlprefix\url{https://linkinghub.elsevier.com/retrieve/pii/S0191261516303319}

\bibitem{Lee2017a}
C.~Lee, J.~Han, \href{https://linkinghub.elsevier.com/retrieve/pii/S0191261517305052}{{Benders-and-Price approach for electric vehicle charging station location problem under probabilistic travel range}}, Transportation Research Part B: Methodological 106 (2017) 130--152.
\newblock \href {https://doi.org/10.1016/j.trb.2017.10.011} {\path{doi:10.1016/j.trb.2017.10.011}}.
\newline\urlprefix\url{https://linkinghub.elsevier.com/retrieve/pii/S0191261517305052}

\bibitem{Zhang2017a}
A.~Zhang, J.~E. Kang, C.~Kwon, \href{https://linkinghub.elsevier.com/retrieve/pii/S0191261516304349}{{Incorporating demand dynamics in multi-period capacitated fast-charging location planning for electric vehicles}}, Transportation Research Part B: Methodological 103 (2017) 5--29.
\newblock \href {https://doi.org/10.1016/j.trb.2017.04.016} {\path{doi:10.1016/j.trb.2017.04.016}}.
\newline\urlprefix\url{https://linkinghub.elsevier.com/retrieve/pii/S0191261516304349}

\bibitem{Yldz2019}
B.~Yıldız, E.~Olcaytu, A.~Şen, \href{https://linkinghub.elsevier.com/retrieve/pii/S0191261517311402}{{The urban recharging infrastructure design problem with stochastic demands and capacitated charging stations}}, Transportation Research Part B: Methodological 119 (2019) 22--44.
\newblock \href {https://doi.org/10.1016/j.trb.2018.11.001} {\path{doi:10.1016/j.trb.2018.11.001}}.
\newline\urlprefix\url{https://linkinghub.elsevier.com/retrieve/pii/S0191261517311402}

\bibitem{Xu2020}
M.~Xu, Q.~Meng, \href{https://linkinghub.elsevier.com/retrieve/pii/S0191261519306976}{{Optimal deployment of charging stations considering path deviation and nonlinear elastic demand}}, Transportation Research Part B: Methodological 135 (2020) 120--142.
\newblock \href {https://doi.org/10.1016/j.trb.2020.03.001} {\path{doi:10.1016/j.trb.2020.03.001}}.
\newline\urlprefix\url{https://linkinghub.elsevier.com/retrieve/pii/S0191261519306976}

\bibitem{Lorentzen2017}
E.~Lorentzen, P.~Haugneland, C.~Bu, E.~Hauge, {Charging infrastructure experiences in Norway – the worlds most advanced EV market}, in: EVS30 Symposium, 2017, pp. 1--11.

\bibitem{Yuan2017}
W.~Yuan, J.~Huang, Y.~J. Zhang, \href{http://ieeexplore.ieee.org/document/7352372/}{{Competitive Charging Station Pricing for Plug-In Electric Vehicles}}, IEEE Transactions on Smart Grid 8~(2) (2015) 1--13.
\newblock \href {https://doi.org/10.1109/TSG.2015.2504502} {\path{doi:10.1109/TSG.2015.2504502}}.
\newline\urlprefix\url{http://ieeexplore.ieee.org/document/7352372/}

\bibitem{Dong2018}
X.~Dong, Y.~Mu, X.~Xu, H.~Jia, J.~Wu, X.~Yu, Y.~Qi, \href{https://doi.org/10.1016/j.apenergy.2018.05.042 https://linkinghub.elsevier.com/retrieve/pii/S0306261918307359}{{A charging pricing strategy of electric vehicle fast charging stations for the voltage control of electricity distribution networks}}, Applied Energy 225~(92) (2018) 857--868.
\newblock \href {https://doi.org/10.1016/j.apenergy.2018.05.042} {\path{doi:10.1016/j.apenergy.2018.05.042}}.
\newline\urlprefix\url{https://doi.org/10.1016/j.apenergy.2018.05.042 https://linkinghub.elsevier.com/retrieve/pii/S0306261918307359}

\bibitem{Wang2018a}
S.~Wang, S.~Bi, Y.~J.~A. Zhang, J.~Huang, {Electrical Vehicle Charging Station Profit Maximization: Admission, Pricing, and Online Scheduling}, IEEE Transactions on Sustainable Energy 9~(4) (2018) 1722--1731.
\newblock \href {https://doi.org/10.1109/TSTE.2018.2810274} {\path{doi:10.1109/TSTE.2018.2810274}}.

\bibitem{Hu2016}
Z.~Hu, K.~Zhan, H.~Zhang, Y.~Song, \href{http://dx.doi.org/10.1016/j.apenergy.2016.06.025}{{Pricing mechanisms design for guiding electric vehicle charging to fill load valley}}, Applied Energy 178 (2016) 155--163.
\newblock \href {https://doi.org/10.1016/j.apenergy.2016.06.025} {\path{doi:10.1016/j.apenergy.2016.06.025}}.
\newline\urlprefix\url{http://dx.doi.org/10.1016/j.apenergy.2016.06.025}

\bibitem{Bayram2015}
I.~S. Bayram, G.~Michailidis, M.~Devetsikiotis, \href{http://ieeexplore.ieee.org/document/6940323/}{{Unsplittable Load Balancing in a Network of Charging Stations Under QoS Guarantees}}, IEEE Transactions on Smart Grid 6~(3) (2015) 1292--1302.
\newblock \href {https://doi.org/10.1109/TSG.2014.2362994} {\path{doi:10.1109/TSG.2014.2362994}}.
\newline\urlprefix\url{http://ieeexplore.ieee.org/document/6940323/}

\bibitem{Ban2012}
D.~Ban, G.~Michailidis, M.~Devetsikiotis, \href{http://ieeexplore.ieee.org/document/6175601/}{{Demand response control for PHEV charging stations by dynamic price adjustments}}, in: 2012 IEEE PES Innovative Smart Grid Technologies (ISGT), IEEE, 2012, pp. 1--8.
\newblock \href {https://doi.org/10.1109/ISGT.2012.6175601} {\path{doi:10.1109/ISGT.2012.6175601}}.
\newline\urlprefix\url{http://ieeexplore.ieee.org/document/6175601/}

\bibitem{Escudero-Garzas2012}
J.~J. Escudero-Garzas, G.~Seco-Granados, \href{http://ieeexplore.ieee.org/document/6175791/}{{Charging station selection optimization for plug-in electric vehicles: An oligopolistic game-theoretic framework}}, in: 2012 IEEE PES Innovative Smart Grid Technologies (ISGT), IEEE, 2012, pp. 1--8.
\newblock \href {https://doi.org/10.1109/ISGT.2012.6175791} {\path{doi:10.1109/ISGT.2012.6175791}}.
\newline\urlprefix\url{http://ieeexplore.ieee.org/document/6175791/}

\bibitem{Lee2015}
W.~Lee, L.~Xiang, R.~Schober, V.~W.~S. Wong, \href{http://ieeexplore.ieee.org/document/6987327/}{{Electric Vehicle Charging Stations With Renewable Power Generators: A Game Theoretical Analysis}}, IEEE Transactions on Smart Grid 6~(2) (2015) 608--617.
\newblock \href {https://doi.org/10.1109/TSG.2014.2374592} {\path{doi:10.1109/TSG.2014.2374592}}.
\newline\urlprefix\url{http://ieeexplore.ieee.org/document/6987327/}

\bibitem{Moghaddam2019}
Z.~Moghaddam, I.~Ahmad, D.~Habibi, M.~A.~S. Masoum, \href{https://ieeexplore.ieee.org/document/8632961/}{{A Coordinated Dynamic Pricing Model for Electric Vehicle Charging Stations}}, IEEE Transactions on Transportation Electrification 5~(1) (2019) 226--238.
\newblock \href {https://doi.org/10.1109/TTE.2019.2897087} {\path{doi:10.1109/TTE.2019.2897087}}.
\newline\urlprefix\url{https://ieeexplore.ieee.org/document/8632961/}

\bibitem{LATINOPOULOS2017175}
C.~Latinopoulos, A.~Sivakumar, J.~Polak, \href{https://www.sciencedirect.com/science/article/pii/S0968090X17301134}{Response of electric vehicle drivers to dynamic pricing of parking and charging services: Risky choice in early reservations}, Transportation Research Part C: Emerging Technologies 80 (2017) 175--189.
\newblock \href {https://doi.org/https://doi.org/10.1016/j.trc.2017.04.008} {\path{doi:https://doi.org/10.1016/j.trc.2017.04.008}}.
\newline\urlprefix\url{https://www.sciencedirect.com/science/article/pii/S0968090X17301134}

\bibitem{Wu2017}
F.~Wu, R.~Sioshansi, \href{https://linkinghub.elsevier.com/retrieve/pii/S0191261516305343}{{A two-stage stochastic optimization model for scheduling electric vehicle charging loads to relieve distribution-system constraints}}, Transportation Research Part B: Methodological 102 (2017) 55--82.
\newblock \href {https://doi.org/10.1016/j.trb.2017.05.002} {\path{doi:10.1016/j.trb.2017.05.002}}.
\newline\urlprefix\url{https://linkinghub.elsevier.com/retrieve/pii/S0191261516305343}

\bibitem{YANG2021103186}
D.~Yang, N.~J. Sarma, M.~F. Hyland, R.~Jayakrishnan, \href{https://www.sciencedirect.com/science/article/pii/S0968090X21002023}{Dynamic modeling and real-time management of a system of ev fast-charging stations}, Transportation Research Part C: Emerging Technologies 128 (2021) 103186.
\newblock \href {https://doi.org/https://doi.org/10.1016/j.trc.2021.103186} {\path{doi:https://doi.org/10.1016/j.trc.2021.103186}}.
\newline\urlprefix\url{https://www.sciencedirect.com/science/article/pii/S0968090X21002023}

\bibitem{Mourad2019}
A.~Mourad, J.~Puchinger, C.~Chu, \href{https://linkinghub.elsevier.com/retrieve/pii/S0191261518304776}{{A survey of models and algorithms for optimizing shared mobility}}, Transportation Research Part B: Methodological 123 (2019) 323--346.
\newblock \href {https://doi.org/10.1016/j.trb.2019.02.003} {\path{doi:10.1016/j.trb.2019.02.003}}.
\newline\urlprefix\url{https://linkinghub.elsevier.com/retrieve/pii/S0191261518304776}

\bibitem{Wang2019a}
H.~Wang, H.~Yang, \href{https://linkinghub.elsevier.com/retrieve/pii/S019126151831172X}{{Ridesourcing systems: A framework and review}}, Transportation Research Part B: Methodological 129 (2019) 122--155.
\newblock \href {https://doi.org/10.1016/j.trb.2019.07.009} {\path{doi:10.1016/j.trb.2019.07.009}}.
\newline\urlprefix\url{https://linkinghub.elsevier.com/retrieve/pii/S019126151831172X}

\bibitem{Buckley2018}
L.~Buckley, S.-A. Kaye, A.~K. Pradhan, \href{https://linkinghub.elsevier.com/retrieve/pii/S1369847817305168}{{A qualitative examination of drivers' responses to partially automated vehicles}}, Transportation Research Part F: Traffic Psychology and Behaviour 56 (2018) 167--175.
\newblock \href {https://doi.org/10.1016/j.trf.2018.04.012} {\path{doi:10.1016/j.trf.2018.04.012}}.
\newline\urlprefix\url{https://linkinghub.elsevier.com/retrieve/pii/S1369847817305168}

\bibitem{Kyriakidis2015}
M.~Kyriakidis, R.~Happee, J.~de~Winter, \href{https://linkinghub.elsevier.com/retrieve/pii/S1369847815000777}{{Public opinion on automated driving: Results of an international questionnaire among 5000 respondents}}, Transportation Research Part F: Traffic Psychology and Behaviour 32 (2015) 127--140.
\newblock \href {https://doi.org/10.1016/j.trf.2015.04.014} {\path{doi:10.1016/j.trf.2015.04.014}}.
\newline\urlprefix\url{https://linkinghub.elsevier.com/retrieve/pii/S1369847815000777}

\bibitem{Fagnant2015}
D.~J. Fagnant, K.~Kockelman, \href{https://linkinghub.elsevier.com/retrieve/pii/S0965856415000804}{{Preparing a nation for autonomous vehicles: opportunities, barriers and policy recommendations}}, Transportation Research Part A: Policy and Practice 77 (2015) 167--181.
\newblock \href {https://doi.org/10.1016/j.tra.2015.04.003} {\path{doi:10.1016/j.tra.2015.04.003}}.
\newline\urlprefix\url{https://linkinghub.elsevier.com/retrieve/pii/S0965856415000804}

\bibitem{Sun2013}
Z.~Sun, B.~Zan, X.~J. Ban, M.~Gruteser, \href{https://linkinghub.elsevier.com/retrieve/pii/S0191261513001252}{{Privacy protection method for fine-grained urban traffic modeling using mobile sensors}}, Transportation Research Part B: Methodological 56 (2013) 50--69.
\newblock \href {https://doi.org/10.1016/j.trb.2013.07.010} {\path{doi:10.1016/j.trb.2013.07.010}}.
\newline\urlprefix\url{https://linkinghub.elsevier.com/retrieve/pii/S0191261513001252}

\bibitem{ismagilova2020security}
E.~Ismagilova, L.~Hughes, N.~P. Rana, Y.~K. Dwivedi, Security, privacy and risks within smart cities: Literature review and development of a smart city interaction framework, Information Systems Frontiers (2020) 1--22.

\bibitem{huang2017secure}
Q.~Huang, L.~Wang, Y.~Yang, Secure and privacy-preserving data sharing and collaboration in mobile healthcare social networks of smart cities, Security and Communication Networks 2017 (2017).

\bibitem{vaidya2021privacy}
G.~Vaidya, P.~Bindra, M.~Kshirsagar, S.~C. Tamane, Privacy and security technologies for smart city development, in: Security and Privacy Applications for Smart City Development, Springer, 2021, pp. 3--23.

\bibitem{BAE2021}
S.~Bae, S.~Gros, B.~Kulcsár, Can ai abuse personal information in an ev fast-charging market?, IEEE Transactions on Intelligent Transportation Systems 23~(7) (2022) 8759--8769.
\newblock \href {https://doi.org/10.1109/TITS.2021.3086006} {\path{doi:10.1109/TITS.2021.3086006}}.

\bibitem{Sutton2018}
R.~S. Sutton, A.~G. Barto, {Reinforcement learning: an introduction}, MIT Press, 2018.

\bibitem{teslasemi2020}
\href{https://www.tesla.com/semi?redirect=no}{{Tesla Semi}} (2020).
\newline\urlprefix\url{https://www.tesla.com/semi?redirect=no}

\bibitem{Darmann2012a}
A.~Darmann, E.~Elkind, S.~Kurz, J.~Lang, J.~Schauer, G.~Woeginger, \href{http://link.springer.com/10.1007/978-3-642-35311-6{\_}12 http://link.springer.com/10.1007/978-3-642-35311-6}{{Internet and Network Economics}}, Vol. 7695 of Lecture Notes in Computer Science, Springer Berlin Heidelberg, Berlin, Heidelberg, 2012.
\newblock \href {https://doi.org/10.1007/978-3-642-35311-6} {\path{doi:10.1007/978-3-642-35311-6}}.
\newline\urlprefix\url{http://link.springer.com/10.1007/978-3-642-35311-6{\_}12 http://link.springer.com/10.1007/978-3-642-35311-6}

\bibitem{Darmann2018}
A.~Darmann, E.~Elkind, S.~Kurz, J.~Lang, J.~Schauer, G.~Woeginger, \href{http://link.springer.com/10.1007/s00182-017-0596-4}{{Group activity selection problem with approval preferences}}, International Journal of Game Theory 47~(3) (2018) 767--796.
\newblock \href {https://doi.org/10.1007/s00182-017-0596-4} {\path{doi:10.1007/s00182-017-0596-4}}.
\newline\urlprefix\url{http://link.springer.com/10.1007/s00182-017-0596-4}

\bibitem{Gunnel2023}
G.~B{\aa}ngman, \href{https://bransch.trafikverket.se/contentassets/4b1c1005597d47bda386d81dd3444b24/2023/19_english_summary_a712.pdf}{{Analysmetod och samh{\"{a}}llsekonomiska kalkylv{\"{a}}rden f{\"{o}}r transportsektorn: ASEK 6.1}}, Tech. rep., Trafikverket (2023).
\newline\urlprefix\url{https://bransch.trafikverket.se/contentassets/4b1c1005597d47bda386d81dd3444b24/2023/19_english_summary_a712.pdf}

\bibitem{Yuan2017a}
W.~Yuan, J.~Huang, Y.~J. Zhang, {Competitive Charging Station Pricing for Plug-In Electric Vehicles}, IEEE Transactions on Smart Grid 8~(2) (2017) 627--639.
\newblock \href {http://arxiv.org/abs/1511.07907} {\path{arXiv:1511.07907}}, \href {https://doi.org/10.1109/TSG.2015.2504502} {\path{doi:10.1109/TSG.2015.2504502}}.

\bibitem{Bhattacharya2016}
S.~Bhattacharya, K.~Kar, J.~H. Chow, A.~Gupta, \href{http://ieeexplore.ieee.org/document/7440866/}{{Extended Second Price Auctions With Elastic Supply for PEV Charging in the Smart Grid}}, IEEE Transactions on Smart Grid 7~(4) (2016) 2082--2093.
\newblock \href {https://doi.org/10.1109/TSG.2016.2546281} {\path{doi:10.1109/TSG.2016.2546281}}.
\newline\urlprefix\url{http://ieeexplore.ieee.org/document/7440866/}

\bibitem{Li2011b}
N.~Li, L.~Chen, S.~H. Low, \href{https://ieeexplore.ieee.org/document/6039082/}{{Optimal demand response based on utility maximization in power networks}}, in: 2011 IEEE Power and Energy Society General Meeting, IEEE, 2011, pp. 1--8.
\newblock \href {https://doi.org/10.1109/PES.2011.6039082} {\path{doi:10.1109/PES.2011.6039082}}.
\newline\urlprefix\url{https://ieeexplore.ieee.org/document/6039082/}

\bibitem{Bitar2013}
E.~Bitar, {Yunjian Xu}, \href{http://ieeexplore.ieee.org/document/6760775/}{{On incentive compatibility of deadline differentiated pricing for deferrable demand}}, in: 52nd IEEE Conference on Decision and Control, IEEE, 2013, pp. 5620--5627.
\newblock \href {https://doi.org/10.1109/CDC.2013.6760775} {\path{doi:10.1109/CDC.2013.6760775}}.
\newline\urlprefix\url{http://ieeexplore.ieee.org/document/6760775/}

\bibitem{Gharesifard2013}
B.~Gharesifard, T.~Basar, A.~D. Dominguez-Garcia, \href{http://ieeexplore.ieee.org/document/6580628/}{{Price-based distributed control for networked plug-in electric vehicles}}, in: 2013 American Control Conference, IEEE, 2013, pp. 5086--5091.
\newblock \href {https://doi.org/10.1109/ACC.2013.6580628} {\path{doi:10.1109/ACC.2013.6580628}}.
\newline\urlprefix\url{http://ieeexplore.ieee.org/document/6580628/}

\bibitem{Wang2016}
Q.~Wang, X.~Liu, J.~Du, F.~Kong, \href{http://ieeexplore.ieee.org/document/7383228/}{{Smart Charging for Electric Vehicles: A Survey From the Algorithmic Perspective}}, IEEE Communications Surveys {\&} Tutorials 18~(2) (2016) 1500--1517.
\newblock \href {https://doi.org/10.1109/COMST.2016.2518628} {\path{doi:10.1109/COMST.2016.2518628}}.
\newline\urlprefix\url{http://ieeexplore.ieee.org/document/7383228/}

\bibitem{Jang2018g}
I.~Jang, H.~S. Shin, A.~Tsourdos, {Anonymous Hedonic Game for Task Allocation in a Large-Scale Multiple Agent System}, IEEE Transactions on Robotics (2018) 1--15\href {http://arxiv.org/abs/1711.06871} {\path{arXiv:1711.06871}}, \href {https://doi.org/10.1109/TRO.2018.2858292} {\path{doi:10.1109/TRO.2018.2858292}}.

\bibitem{YING2020210}
C.~shuo Ying, A.~H. Chow, K.-S. Chin, \href{https://www.sciencedirect.com/science/article/pii/S0191261520303829}{An actor-critic deep reinforcement learning approach for metro train scheduling with rolling stock circulation under stochastic demand}, Transportation Research Part B: Methodological 140 (2020) 210--235.
\newblock \href {https://doi.org/https://doi.org/10.1016/j.trb.2020.08.005} {\path{doi:https://doi.org/10.1016/j.trb.2020.08.005}}.
\newline\urlprefix\url{https://www.sciencedirect.com/science/article/pii/S0191261520303829}

\bibitem{AHAMED2021227}
T.~Ahamed, B.~Zou, N.~P. Farazi, T.~Tulabandhula, \href{https://www.sciencedirect.com/science/article/pii/S0191261521001636}{Deep reinforcement learning for crowdsourced urban delivery}, Transportation Research Part B: Methodological 152 (2021) 227--257.
\newblock \href {https://doi.org/https://doi.org/10.1016/j.trb.2021.08.015} {\path{doi:https://doi.org/10.1016/j.trb.2021.08.015}}.
\newline\urlprefix\url{https://www.sciencedirect.com/science/article/pii/S0191261521001636}

\bibitem{goodfellow2016deep}
I.~Goodfellow, Y.~Bengio, A.~Courville, \href{https://books.google.co.in/books?id=Np9SDQAAQBAJ}{Deep Learning}, Adaptive computation and machine learning, MIT Press, 2016.
\newline\urlprefix\url{https://books.google.co.in/books?id=Np9SDQAAQBAJ}

\bibitem{Mnih2015}
V.~Mnih, K.~Kavukcuoglu, D.~Silver, A.~A. Rusu, J.~Veness, M.~G. Bellemare, A.~Graves, M.~Riedmiller, A.~K. Fidjeland, G.~Ostrovski, S.~Petersen, C.~Beattie, A.~Sadik, I.~Antonoglou, H.~King, D.~Kumaran, D.~Wierstra, S.~Legg, D.~Hassabis, \href{http://www.nature.com/articles/nature14236}{{Human-level control through deep reinforcement learning}}, Nature 518~(7540) (2015) 529--533.
\newblock \href {https://doi.org/10.1038/nature14236} {\path{doi:10.1038/nature14236}}.
\newline\urlprefix\url{http://www.nature.com/articles/nature14236}

\bibitem{Bae2020}
S.~Bae, I.~Jang, S.~Gros, B.~Kulcsár, J.~Hellgren, A game approach for charging station placement based on user preferences and crowdedness, IEEE Transactions on Intelligent Transportation Systems (2020) 1--16\href {https://doi.org/10.1109/TITS.2020.3038938} {\path{doi:10.1109/TITS.2020.3038938}}.

\bibitem{yadati2009}
Y.~Narahari, R.~Narayanam, D.~Grag, H.~Prakash, \href{http://link.springer.com/10.1007/978-1-84800-938-7}{{Game Theoretic Problems in Network Economics and Mechanism Design Solutions}}, Springer London, London, 2009.
\newblock \href {https://doi.org/10.1007/978-1-84800-938-7} {\path{doi:10.1007/978-1-84800-938-7}}.
\newline\urlprefix\url{http://link.springer.com/10.1007/978-1-84800-938-7}

\end{thebibliography}

\end{document}